\documentclass[10pt, twocolumn, twoside]{IEEEtran}

\usepackage{cite}

\usepackage{graphicx,amssymb, dsfont, setspace, amsfonts,amsthm,color}
\usepackage[cmex10]{amsmath}

\DeclareMathOperator{\e}{e}
\DeclareMathOperator{\Pro}{Pr}

\newtheorem{lemma}{Lemma}
\newtheorem{theorem}{Theorem}
\newtheorem{defin}{Definition}

\ifCLASSINFOpdf
\else
\fi

\begin{document}

\title{Poisson Group Testing: A Probabilistic Model for Boolean Compressed Sensing}

\author{Amin~Emad$^\ast$,~\IEEEmembership{Student~Member,~IEEE,}
       and Olgica~Milenkovic,~\IEEEmembership{Senior Member,~IEEE}%
       \vspace{-15pt}
\thanks{The authors are with the Department of Electrical and Computer Engineering, University of Illinois at Urbana-Champaign, Urbana,
IL, 61801 USA e-mail: \{emad2,milenkov\}@illinois.edu. This work was supported in parts by NSF grants CCF 0809895, CCF 1218764 and the Emerging Frontiers for Science of Information Center, CCF 0939370. This paper was presented, in part, at the International Conference on Acoustics, Speech, and Signal Processing (ICASSP), 2014~\cite{AM14}.}}

\maketitle
\begin{abstract}
We introduce a novel probabilistic group testing framework, termed Poisson group testing, in which the number of defectives follows a right-truncated Poisson distribution. The Poisson model has a number of new applications, including dynamic testing with diminishing relative rates of defectives. We consider both nonadaptive and semi-adaptive identification methods. For nonadaptive methods, we derive a lower bound on the number of tests required to identify the defectives with a probability of error that asymptotically converges to zero; in addition, we propose test matrix constructions for which the number of tests closely matches the lower bound. For semi-adaptive methods, we describe a lower bound on the expected number of tests required to identify the defectives with zero error probability. In addition, we propose a stage-wise reconstruction algorithm for which the expected number of tests is only a constant factor away from the lower bound. The methods rely only on an estimate of the average number of defectives, rather than on the individual probabilities of subjects being defective.
 \end{abstract}


\IEEEpeerreviewmaketitle

\vspace*{-10pt}
\section{Introduction}\label{sec:intro}

Group testing (GT), also known as Boolean compressed sensing, is a method for identifying a group of subjects with some distinguishable characteristic, frequently referred to as defectives, from a large group of entities~\cite{DH06},~\cite{D04}. The gist of the GT approach is that for a small number of defectives, one can reduce the required number of experiments by testing subgroups of subjects rather than all individuals separately. Given its simple working principles and the potential for reducing the cost of component screening, GT has found many applications in areas as diverse as communication theory, signal processing, bioinformatics, mathematics, and machine learning \cite{W85,A94,FDH95,BBK03,DH00,MV13}.

The test model of the GT framework varies depending on the application at hand. The original setup, also known as conventional GT, was proposed by Dorfman~\cite{D43}, and involves logical OR computations on the test signatures. More precisely, in conventional GT, the result of a test is positive if there exists at least one defective in the test pool and negative otherwise. Many other models have been proposed in the literature, such as the adder channel, also known as quantitative GT~\cite{DH00}, threshold GT~\cite{D06}, and symmetric GT~\cite{AM110}. More recent developments include the semi-quantitative group testing (SQGT) paradigm, which provides a unifying framework for a number of GT models and generalizes the notion of GT to non-binary test matrices and non-binary test outcomes~\cite{AM11},~\cite{AM13}. 
In addition, GT is closely related to compressed sensing (CS)~\cite{CRT06},~\cite{DO06} and integer CS~\cite{DM09}; the main differences lie in the structure of the alphabet used ($\mathbb{R}$ or $\mathbb{C}$ in CS, $\{{0,1\}}$ or a discrete set of integers for GT, and a bounded set of integers for integer CS) and the operations used to perform dimensionality reduction (addition in CS and integer CS, Boolean OR in conventional GT). 

The group testing literature may be divided into two categories based on how the number of defectives is modeled. In combinatorial GT, the number of defectives, or an upper bound on the number of defectives, is fixed and assumed to be known in advance~\cite{DH00}. On the other hand, in probabilistic GT (PGT), the number of defectives is a random variable with a given probability distribution~\cite{D43}. With almost no exceptions, the PGT literature focuses on a Binomial$(n,p_0)$ distribution for the number of defectives. Such a model arises when each of the $n$ subjects is defective with a fixed probability $0<p_0<1$, independent of all other subjects. Binomial models are not necessarily sparse, given that $p_0$ may be a constant and given that the defective selection process is random.  

Here, we propose a novel GT paradigm, termed \emph{Poisson PGT}, which models the distribution of the number of defectives via a right-truncated Poisson distribution with parameter $\lambda(n)=o(n)$. Our motivation for this assumption comes from clinical testing, where one is interested in identifying infected individuals under the assumption that infections gradually die out. A similar scenario is encountered in screening DNA clones for the presence of certain DNA substrings, where the clones are test subjects and defectives are clones that contain the given substrings. The distribution of clones containing a given DNA pattern is frequently modeled as Poisson~\cite{DH00}. Other applications include testing genetic traits that are negatively selected for (i.e., traits that diminish in time, as they reduce the fitness of a species). The assumption $\lambda(n)=o(n)$ ensures that the longer the waiting time or the larger the number of test subjects, the smaller the average \emph{relative} fraction of defectives. In other words, the rate of defectives diminishes with time.

The Poisson PGT model has a number of useful properties that make it an important alternative to classical binomial models. Although a binomial distribution with $p_0\!\ll\!1$ and a large $n$, where $\lambda\!=\!np_0$ is a constant, converges to a Poisson distribution with parameter $\lambda\!=\!np_0$~\cite{P02}, our model allows the parameter $\lambda(n)$ of the (truncated) Poisson distribution to grow with $n$; more precisely, the model and the results derived in this paper are valid even if $\lim_{n\rightarrow\infty}\lambda(n)\!=\!\infty$, as long as $\lim_{n\rightarrow\infty}\frac{\lambda(n)}{n}\!=\!0$. Such a model is useful in settings were test subjects are assumed to arrive sequentially in time, and where tests are performed only once a sufficient number of subjects $n$ is present. This model is also applicable to streaming and dynamic testing scenarios~\cite{GS06}, in which the probability that a subject is defective decreases in time so that newly arriving subjects are less likely to be defective. In such a setting, classical binomial$(n,p_0)$ models are inadequate, as they assume that the probability $p_0$ of a subject being defective does not depend on the number of test subjects. 

A number of papers have considered a Poisson model to capture the streaming dynamics of the \emph{arrivals of subjects} to a test center~\cite{BPPSdDS07},~\cite{CWLB10}. In contrast, our model does not make any assumptions on the distribution of the general subject population, but instead focuses on modeling the number of defectives using a right-truncated Poisson distribution. In addition, the focus of ~\cite{BPPSdDS07},~\cite{CWLB10} is on determining the total amount of time (delay) required to test a batch of subjects arriving at random times. However, here we concentrate on the completely unrelated problem of finding necessary and sufficient conditions on the smallest number of tests needed for accurate nonadaptive and semi-adaptive GT.

In addition, a number of papers have considered the problem of binomial group testing with different subjects having different probabilities of being defective. This line of work was introduced in~\cite{H75} under the name \emph{generalized binomial group testing} (GBGT). Recently, this problem has received renewed interest under the name of \emph{heterogeneous binomial group testing}~\cite{BBT12}. 
In~\cite{H75}, a two-stage algorithm for GBGT was proposed, resulting in a complicated minimization problem for the expected number of tests required; unfortunately, no closed-form expression, nor any simply calculable expression, was provided for the expected number of tests. In~\cite{GH74}, a similar problem was considered in which the goal was to isolate a \emph{single} defective in the GBGT model. For this problem, the authors proposed an optimal adaptive procedure using a binary testing tree, which was obtained for a set of weights that depend on the probabilities of the subjects being defective. In addition, an upper bound on the expected number of tests was provided in the form of a complicated sum. Other papers that consider the GBGT model include~\cite{BBT12},~\cite{KS88}, and \cite{MTB12}. As we explain in the next section, although related to our Poisson model through Le Cam's theorem, GBGT operates under very different prior knowledge assumptions and cannot be considered within the same analytical framework.

The main contributions of this work are three-fold. First, we introduce a novel probabilistic GT model with applications in streaming and dynamic testing scenarios. This model generalizes probabilistic group testing models beyond the binomial GT paradigm with constant $p_0$ and other models previously considered in the literature. Second, we bridge the gap between combinatorial GT and probabilistic GT methodology by showing how the algorithms and analytical tools developed for combinatorial GT can be generalized and adapted for probabilistic GT. To the best of our knowledge, this is the first attempt to analyze combinatorial GT and probabilistic GT within the same framework. Finally, we derive closely matching lower and upper bounds on the number of tests required for finding the defectives in Poisson testing using both nonadaptive and semi-adaptive algorithms.

The paper is organized as follows. Section~\ref{Sec:model} introduces the Poisson GT model, while Sections~\ref{sec:nonadaptive} and~\ref{sec:adaptive} describe the main results of the paper. A summary of the results and an accompanying discussion are provided in Section~\ref{discussion}. In Section~\ref{sec:nonadaptive}, we first use an adaptation of Fano's inequality to find a lower bound on the number of nonadaptive tests required to identify defectives under the Poisson PGT model, with a probability of error converging to zero as the number of subjects grows. We then proceed to describe a simple nonadaptive method based on binary disjunct matrices~\cite{KS64}. The test matrix is constructed probabilistically, with the entries of the matrix being independent and Bernoulli distributed. Given that the number of tests obtained via this method does not tightly match the lower bound, we describe an alternative nonadaptive method with a number of tests differing from the lower bound by only an arbitrary slowly-growing function in $n$. The test matrix in this method does not rely on the disjunctness property and the entries of the matrix are not i.i.d. distributed. In Appendix~\ref{sec:method3}, we use information-theoretic arguments to derive a tight upper bound on the number of nonadaptive tests for Poisson PGT. Following the practice of Binomial probabilistic group testing, in Section~\ref{sec:adaptive} we use Huffman coding to find a lower bound on the \emph{expected number of tests} required by adaptive and semi-adaptive methods to identify the defectives with zero error probability. Then, we show that a simple semi-adaptive algorithm identifies all the defectives with an expected number of tests only a constant factor away from the lower bound.

\vspace*{-5pt}
\section{Problem setup}\label{Sec:model}

Throughout the paper we adopt the following notation. Bold-face uppercase and bold-face lowercase letters denote matrices and vectors, respectively. Simple uppercase letters are used to denote random matrices, random vectors, and random variables; similarly, simple lowercase letters are used for scalars. Calligraphic letters are used to denote sets. For simplicity, we often write $\mathcal{X}=\{\mathbf{x}_i\}_1^s$ for a set of $s$ codewords, $\mathcal{X}=\{\mathbf{x}_1,\mathbf{x}_2,\dots,\mathbf{x}_s\}$. The symbols $\log(\cdot)$ and $\log_2(\cdot)$ are used to denote the natural logarithm and the base $2$ logarithm, respectively. For a \emph{finite} integer $K\geq1$, we also make use of the $K$-fold logarithm function, defined as 
\vspace*{-4pt}
\begin{equation}\label{eq:logk}
\log^{(K)}n\triangleq\underbrace{\log\log\cdots\log}_{K  \  \textnormal{times}} n.
\vspace*{-4pt}
\end{equation}
Note that for $K>1$, this function grows slower than $\log(n)$.

Asymptotic symbols such as $o(\cdot)$ and $O(\cdot)$ are used in the standard manner. More precisely, we say that $f(x) = O(g(x))$ if and only if there exist $M,x_0\in\mathbb{R}$, with $M>0$, such that $|f(x)|\leq M|g(x)|$ for all $x\geq x_0$. Also, $f(x)=o(g(x))$ means that for any $\epsilon>0$, there exists $x_0\in\mathbb{R}$ such that  $|f(x)|\leq \epsilon|g(x)|$ for all $x\geq x_0$.

Let $\mathcal{S}$ denote the set of test subjects with cardinality $n$, among which a subset of $\mathcal{D}$ subjects is defective. In the Poisson PGT model, we assume that the number of defectives follows a right-truncated Poisson distribution with parameters $\lambda(n)$ and $n$, i.e.,
\vspace*{-4pt}
\begin{equation}\label{trunpoisson}
P_D(d)=
\begin{cases}
c(n)\frac{{\lambda(n)}^d}{d!}\e^{-\lambda(n)}, &0\leq d\leq n\\
0,&\textnormal{otherwise.}
\end{cases}.
\vspace*{-4pt}
\end{equation}
Here, $D=|\mathcal{D}|$ denotes the number of defectives and $c(n)=\e^{\lambda(n)}/\sum_{d=0}^n\frac{\lambda(n)^d}{d!}$ is a normalization coefficient. Note that $c(n)$ is a decreasing function of $n$, such that $\lim_{n\rightarrow\infty}c(n)=1$. In addition we assume that all subsets of $\mathcal{S}$ with equal cardinality have the same probability of being defective. This assumption is used to model the setup in which given $D=d$, the decoder has no information as to which set of cardinality $d$ is most likely to be the set of defectives. 

Let $\bar{\lambda}(n)$ be the expected number of defectives in the model. It can be easily verified that 
\vspace*{-4pt}
\begin{align}\nonumber
\bar{\lambda}(n)=\mathbb{E}[D]&=\sum_{d=0}^n c(n)d\frac{{\lambda(n)}^d}{d!}e^{-\lambda(n)}\\\nonumber
&=\lambda\sum_{d=1}^n c(n)\frac{{\lambda(n)}^{d-1}}{(d-1)!}e^{-\lambda(n)}\\\nonumber
&=\lambda\sum_{d=0}^{n-1} c(n)\frac{{\lambda(n)}^{d}}{d!}e^{-\lambda(n)}\\\nonumber
&=\lambda\left(\sum_{d=0}^{n} c(n)\frac{{\lambda(n)}^{d}}{d!}e^{-\lambda(n)}-c(n)\frac{\lambda^n}{n!}e^{-\lambda}\right)\\\label{eq:mean0}
&=\lambda(n)\left(\!1\!-c(n)\!\frac{\lambda(n)^n}{n!}e^{-\lambda(n)}\!\right)\\\nonumber
&=\lambda(n)(1\!-\!o(1)).
\vspace*{-4pt}
\end{align}

A right-truncated Poisson distribution is closely related to a finite support version of the non-uniform Bernoulli model on the set of test subjects, in which the $i^{\text{th}}$ subject is defective with probability $p_i$, $0 \leq p_i \leq 1$, independent of all other test subjects. From Le Cam's theorem~\cite{L60}, it may be deduced that the number of defectives $D$ under this model satisfies
\vspace*{-4pt}
\begin{equation}\label{harmonic}
\sum_{k=0}^{\infty} \: \left|\Pro({D=d})-\e^{-\lambda(n)} \, \frac{\lambda(n)^d}{d!}\right| \leq 2\sum_{i=1}^{n}p_i^2,
\vspace*{-4pt}
\end{equation}
where $\lambda(n)=\sum_{i}^{n} p_i$. As an example, one can choose $p_i=c/i$, for some constant $c>0$, to arrive at a model where individual subjects have decreasing probabilities in $i$ of being defective, so that $\lambda(n)=O(\log\,n)$. The approximation error with respect to the Poisson distribution scales as $2\, c\, \zeta(2)=c\pi^2/3$, where $\zeta(\cdot)$ denotes the Riemann zeta function. By choosing $c$ sufficiently small, the approximation error can be reduced to any desired \emph{positive} level. Although adaptive and other classes of \emph{non-uniform} Bernoulli models were reported in the literature~\cite{H75},~\cite{GH74},~\cite{L14},~\cite{KJP14} they rely on the \emph{exact knowledge} of each probability $p_i$, $i=1,\ldots,n$. However, even in applications in which a subject is defective independently from all other subjects, estimating each of the $p_i$ values may be prohibitively difficult. In contrast, Poisson PGT only makes use of a single aggregate value of the probabilities, $\lambda(n)$, which is less informative but usually much easier to estimate.

In the GT framework, each test is performed on a subset of the subjects, and the result of a test equals $1$ if at least one defective is present in the test, and $0$ otherwise. The total number of tests is denoted by $m$. 
For non-adaptive PGT, despite the fact that the defectives are chosen randomly, the number of tests is \emph{deterministic}. The question of interest is to find the smallest number of tests that guarantee a probability of detection error that converges to zero asymptotically with the number of subjects $n$. In contrast, adaptive and semi-adaptive algorithms, in which tests are performed sequentially or grouped into different stages with the choice at one stage used to inform the choice at the following stages, call for a \emph{random number of tests}. The goal is then to compute the expected number of tests that allows for \emph{zero error probability}. In this paper, we focus on nonadaptive and semi-adaptive testing schemes, and our main results are summarized and discussed in Section~\ref{discussion}.
%

\vspace*{-5pt}
\section{Nonadaptive methods for Poisson PGT}\label{sec:nonadaptive}

Nonadaptive group testing refers to group testing methods in which all tests are designed simultaneously. In other words, in nonadaptive GT the choice of a test is not allowed to depend on the outcomes of previous tests~\cite{DH00}. The main advantage of nonadaptive methods is that all the tests can be performed in parallel, which is of great practical importance for large-scale problems. A clear disadvantage compared to adaptive methods is the sometime significant increase in the number of tests.

In nonadaptive GT, the assignment of subjects to different tests is usually specified via a binary matrix termed the \emph{test matrix}, $\mathbf{C}\in\{0,1\}^{m\times n}$, where $m$ denotes the number of tests and $n$ denotes the number of subjects.  If $\mathbf{C}(i,j)=1$, for $1\leq i\leq m$ and $1\leq j\leq n$, the $j^{\text{th}}$ subject is present in the $i^{\text{th}}$ test; on the other hand, if $\mathbf{C}(i,j)=0$, than the $j^{\text{th}}$ subject is excluded from the $i^{\text{th}}$ test. Throughout this paper we use the terms ``code'' and ``test matrix'' interchangeably; given this definition, a codeword refers to a column of the test matrix. The test results are captured by a vector $\mathbf{y}\in\{0,1\}^m$, frequently referred to as the vector of test results or syndrome. 

\begin{figure}[tbh]
  \centering
  \centerline{\includegraphics[width=0.35\textwidth]{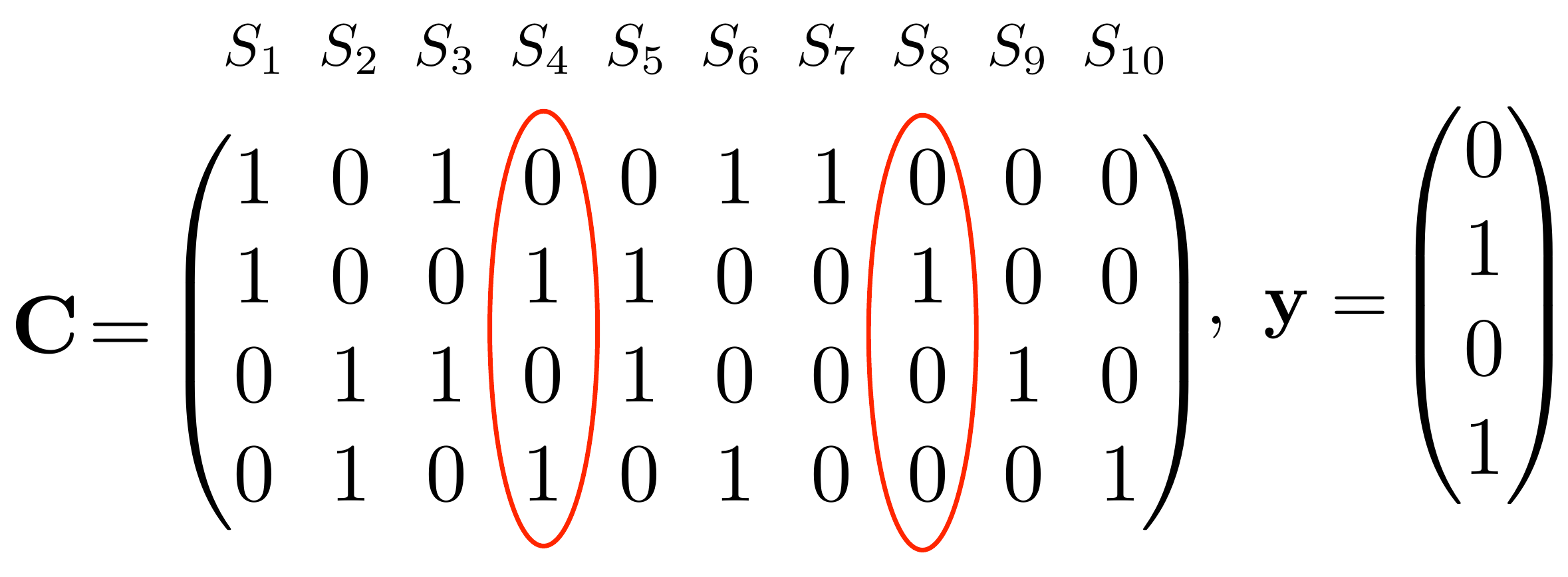}}
\vspace*{-10pt}
 \caption{Example of a test matrix and the test results where $n=10$, $m=4$, and the set of defectives is $\mathcal{D}=\{S_4,S_8\}$.}\label{fig1}
 \vspace*{-10pt}
\end{figure}

It can be easily observed that the vector of test results is equal to the Boolean OR of columns of $\mathbf{C}$ corresponding to the defectives. Fig.~\ref{fig1} illustrates the notion of a test matrix, the set of defectives, and the vector of test results. Note that $S_i$ denotes the $i^{\text{th}}$ subject in $\mathcal{S}$.  
\vspace*{-6pt}

For a fixed test matrix on $n$ subjects, $\mathbf{C}$, and a decoding algorithm $f:(\mathbf{C},\mathbf{y})\mapsto \hat{\mathcal{D}}$, let $\mathcal{E}(n)$ denote the event that the decoding algorithm cannot identify the set of defectives, i.e. the event that $f(\mathbf{C},\mathbf{y})\neq\mathcal{D}$. The ultimate goal of most combinatorial nonadaptive GT methods is to ensure that $P(\mathcal{E}(n))=0$. Due to the probabilistic nature of the Poisson PGT model, any subset of subjects may be defective with a non-zero probability. As a result, since in nonadaptive GT each test is designed independently from previous tests, for any fixed test matrix $\mathbf{C}$ with $m<n$, one can always find a choice of $\mathcal{D}$ for which $f(\mathbf{C},\mathbf{y})\neq\mathcal{D}$. 

To verify the correctness of this claim, consider a fixed test matrix $\mathbf{C}$ and a set of subjects $\mathcal{S}$ such that each column of $\mathbf{C}$ is assigned to one subject in $\mathcal{S}$; for any set $\mathcal{D}'\subseteq\mathcal{S}$, let $\mathbf{y}_{\!\!_{\mathcal{D}'}}$ denote the Boolean OR of the columns of $\mathbf{C}$ corresponding to $\mathcal{D}'$. Since in Poisson PGT each subset of $\mathcal{S}$ may correspond to the set of defectives with a nonzero probability, in order to ensure $P(\mathcal{E}(n))=0$, the test matrix must be able to distinguish between any two distinct subsets of $\mathcal{S}$; in other words, for any two distinct sets $\mathcal{D}_1,\mathcal{D}_2\subseteq\mathcal{S}$, we must have $\mathbf{y}_{\!\!_{\mathcal{D}_1}}\neq\mathbf{y}_{\!\!_{\mathcal{D}_2}}$. Since in total, there exist $\sum_{d=0}^n{n\choose d}=2^n$ choices for the set of defectives, at least $m=n$ tests are required (i.e. one has to test each subject individually).

The discussion above implies that for Poisson PGT, there does not exist a nonadaptive test matrix with fewer than $n$ rows and an accompanying decoding algorithm for which $P(\mathcal{E}(n))=0$; as a result, we instead focus on the requirement that the test matrix satisfy the asymptotic condition 
\vspace*{-4pt}
\begin{equation}\label{condition1}
\lim_{n\rightarrow\infty}P(\mathcal{E}(n))=0.
\vspace*{-4pt}
\end{equation}

In what follows, we propose two nonadaptive test matrix constructions and decoding algorithms to guarantee that the aforementioned condition is met\footnote{Also, we omit the parameter $n$ whenever possible and tacitly assume the dependence of the error probability on this parameter.}. In order to evaluate how effectively each method uses its tests, we first find a lower bound on the minimum number of tests required by a nonadaptive algorithm to ensure~\eqref{condition1}, and then use this bound as a benchmark. The constructive methods provide upper bounds on the minimum number of tests. 

In Section~\ref{sec:lower_bound}, we use Fano's inequality~\cite{CT91} to find a lower bound on the number of tests of the form $(1-\epsilon)\lambda(n)\log_2{n}\  (1-o(1))$, where $\epsilon$ is an arbitrarily small fixed scalar such that $0<\epsilon<1$. In Section~\ref{sec:method1}, we propose a test matrix construction using binary disjunct matrices (Method I). The entries of the test matrix are chosen according to an i.i.d. distribution, and the method requires $m=C_2\kappa(n)\lambda(n)^2\log_2{n}\  (1+o(1))$ measurements, where $\kappa(n)$ is an arbitrary chosen slowly-growing function of $n$. Given the gap between the number of tests in Method I and the lower bound, we propose another method in Section~\ref{sec:method2} that requires only $m=C_1\kappa(n)\lambda(n)\log_2{n}\  (1+o(1))$ tests (Method II). This method is also based on a probabilistic construction, however the entries of the test matrix no longer follow an i.i.d. distribution. One major difference between these two methods is that Method I uses the disjunctness property~\cite{KS64, DH06}, while for Method II we relaxed this constraint. Both of these constructions can be extended to identify the set of defectives in the presence of errors in the vector of test results, and also employ a decoding algorithm with computational complexity $O(mn)$. 

In Appendix~\ref{sec:method3}, we use a standard information theoretic approach -- combined with a maximum likelihood decoder -- to determine an upper bound on the minimum number of tests required by any nonadaptive method based on an i.i.d. test matrix. The number of tests using this approach tightly matches the lower bound under some constraints on the growth of $\lambda(n)$ with respect to $n$. 

 A summary of these results are provided in Section~\ref{discussion}.

\vspace*{-10pt}
\subsection{Lower Bound on the minimum number of tests}\label{sec:lower_bound}

Let $\hat{\mathcal{D}}$ be the set of defectives recovered by some decoding algorithm using a fixed test matrix $\mathbf{C}\in\{0,1\}^{m\times n}$, and given the random vector of test results, $Y\in \{0,1\}^m$. 
Conditioned on $D=d$, $0\leq d\leq n$, let $\mathcal{E}_d$ be the error event that $\hat{\mathcal{D}}\neq\mathcal{D}$; consequently, $P(\mathcal{E})=\mathbb{E}_D[P(\mathcal{E}_d)]$. Using Fano's inequality~\cite{CT91}, one has
\vspace*{-4pt}
\begin{equation}\label{app:temp1}
H(\mathcal{D}|Y,C,D=d)\leq 1+ P(\mathcal{E}_d)\  \log_2{n\choose d},
\vspace*{-4pt}
\end{equation}
where $H(\cdot)$ denotes the Shannon entropy function~\cite{CT91}. 
Since conditioned on $D=d$, the set of defectives $\mathcal{D}$ is chosen uniformly at random, independent on ${\mathbf{C}}$, one has
\vspace*{-4pt}
\begin{equation}\label{app:temp2}
{H(\mathcal{D}|\mathbf{C},D=d)}=\log_2{n\choose d}.
\vspace*{-4pt}
\end{equation}
Using the definition of mutual information~\cite{CT91}, we may write
\vspace*{-4pt}\begin{align}\nonumber
H(\mathcal{D}|\mathbf{C},D)&=H(\mathcal{D}|Y,\mathbf{C},{D})\!+\!I(\mathcal{D};Y|\mathbf{C},D)\\\nonumber
&=H(\mathcal{D}|Y,\mathbf{C},{D})\!+\!H(Y|\mathbf{C},D)\!-\!H(Y|\mathbf{C},\mathcal{D},D)\\\nonumber
&\leq H(\mathcal{D}|Y,\mathbf{C},{D})\!+\!H(Y|D)\!-\!H(Y|\mathbf{C}_\mathcal{D},D)\\\label{app:temp3}
&= H(\mathcal{D}|Y,\mathbf{C},{D})\!+\!I(Y;\mathbf{C}_\mathcal{D}|D),
\vspace*{-4pt}\end{align}
where the inequality follows since conditioning reduces entropy; also, the test results $Y$ only depend on the codewords assigned to the set $\mathcal{D}$ and hence $H(Y|\mathbf{C},\mathcal{D},D)=H(Y|\mathbf{C}_\mathcal{D},D)$, where $\mathbf{C}_\mathcal{D}$ is the set of columns of $\mathbf{C}$ corresponding to $\mathcal{D}$.
Substituting~\eqref{app:temp1} and~\eqref{app:temp2} in~\eqref{app:temp3} yields
\vspace*{-4pt}\begin{align}\nonumber
\log_2{n\choose d}&\leq 1+ P(\mathcal{E}_d)\  \log_2{n\choose d}+I(Y;\mathbf{C}_{\mathcal{D}}|D)\\
\nonumber 
&\Rightarrow P(\mathcal{E}_d)\geq 1-\frac{I(Y;\mathbf{C}_{\mathcal{D}}|D)+1}{\log_2{n\choose d}}.
\vspace*{-4pt}\end{align}
On the other hand, from the following chain of inequalities
\vspace*{-4pt}
\begin{equation}\nonumber
I(Y;\mathbf{C}_{\mathcal{D}}|D)\leq H(Y|D)\leq H(Y)\leq\sum_{i=1}^mH(Y(i))\leq m,
\vspace*{-4pt}
\end{equation}
it follows that
\vspace*{-4pt}
\begin{equation}\label{eq16_0}
P(\mathcal{E}_d)\geq 1-\frac{I(Y;\mathbf{C}_{\mathcal{D}}|D)+1}{\log_2{n\choose d}}\geq 1-\frac{m+1}{\log_2{n\choose d}}.
\vspace*{-4pt}
\end{equation}
Since $P(\mathcal{E})=\mathbb{E}_D[P(\mathcal{E}_d)]$,~\eqref{eq16_0} may be used to find a lower bound on $m$ that ensures $P(\mathcal{E})=o(1)$, as formally stated in the next theorem.
\vspace{-0.1in}

\begin{theorem}\label{theorem:lower}
Let $0<\epsilon<1$ be an arbitrarily small fixed scalar, and suppose that $\lambda(n)=o(n)$. Any nonadaptive group testing method designed for Poisson PGT that satisfies $\lim_{n\rightarrow\infty}P(\mathcal{E})=0$ requires at least $m=(1-\epsilon)\lambda(n)\log_2{n}\  (1-o(1))$ tests.
\end{theorem}
\begin{proof}
Let $0<\epsilon<1$. Then, since $\lambda(n)=o(n)$, for large enough values of $n$, $\lambda(n)(1+\epsilon)<n$. On the other hand, since $P(\mathcal{E}_d)\geq 0$, $0\leq d\leq n$, then 
\vspace*{-4pt}\begin{align}\nonumber
P(\mathcal{E})=\mathbb{E}_D[P(\mathcal{E}_d)]&=\sum_{d=0}^n P(D=d)P(\mathcal{E}_d)\\\nonumber
&\geq\sum_{d=\lambda(n)(1-\epsilon)}^{\lambda(n)(1+\epsilon)} P(D=d)P(\mathcal{E}_d).
\vspace*{-4pt}\end{align}
As a result, a necessary condition for $P(\mathcal{E})=o(1)$ is that for $n$ large enough, 
\vspace*{-4pt}\begin{align}\nonumber
o(1)&\geq \sum_{d=\lambda(n)(1-\epsilon)}^{\lambda(n)(1+\epsilon)} P(D=d)P(\mathcal{E}_d)\\\nonumber
&\geq\sum_{d=\lambda(n)(1-\epsilon)}^{\lambda(n)(1+\epsilon)}P(D=d)\left(1-\frac{m+1}{\log_2{n\choose d}}\right)\\\label{eq16}
&\geq\left[\min_{d:\lambda(1-\epsilon)\leq d\leq\lambda(1+\epsilon)}\left(\!1\!-\!\frac{m+1}{\log_2\!{n\choose d}}\!\right)\right]\!\!\!\sum_{\  d=\lambda(1-\epsilon)}^{\lambda(1+\epsilon)}\hspace{-10pt}P(D=d),
\vspace*{-4pt}\end{align}
where the second inequality is a consequence of Equation~\eqref{eq16_0}.

Using the Chernoff bound for a standard Poisson distribution, it may be shown that
\vspace*{-4pt}
\begin{equation}\nonumber
\sum_{d=\lambda(1+\epsilon)}^{\infty}\hspace{-10pt}\frac{\lambda(n)^d}{d!}\e^{-\lambda(n)}\leq\exp\left(-\lambda(n)(1\!+\!\epsilon)\log(1\!+\!\epsilon)+\epsilon\lambda(n)\right),
\vspace*{-4pt}
\end{equation}
and
\vspace*{-4pt}
\begin{equation}\nonumber
\sum_{d=0}^{\lambda(1\!-\!\epsilon)}\hspace{-3pt}\frac{\lambda(n)^d}{d!}\e^{-\lambda(n)}\leq\exp\left(-\lambda(n)(1-\epsilon)\log(1\!-\!\epsilon)-\epsilon\lambda(n)\right).
\vspace*{-4pt}
\end{equation}
Let $f_1(\epsilon)=(1+\epsilon)\log(1+\epsilon)-\epsilon$ and $f_2(\epsilon)=(1-\epsilon)\log(1-\epsilon)+\epsilon$. It can be easily verified that for $\epsilon>0$, $f_1(\epsilon)>0$ and $f_2(\epsilon)>0$. As a result, from the above inequalities we obtain
\vspace*{-4pt}\begin{equation}\label{eq17}
\sum_{d=\lambda(n)(1-\epsilon)}^{\lambda(n)(1+\epsilon)}P(D=d)\geq c(n)\left(1-\e^{-\lambda(n) f_1(\epsilon)}-\e^{-\lambda(n) f_2(\epsilon)}\right).
\vspace*{-4pt}\end{equation}
On the other hand, for $n>2(1-\epsilon)\lambda(n)$, 
\vspace*{-4pt}\begin{equation}\label{eq18}
\min_{d:\lambda(1-\epsilon)\leq d\leq\lambda(1+\epsilon)}\left(\!1\!-\!\frac{m+1}{\log_2\!{n\choose d}}\!\right)\!=\!\left(\!1\!-\!\frac{m+1}{\log_2\!{n\choose (1-\epsilon)\lambda}}\!\right).
\end{equation}
\vspace*{1pt}

Substituting~\eqref{eq17} and~\eqref{eq18} in~\eqref{eq16}, a necessary condition for $\lim_{n\rightarrow\infty}P(\mathcal{E})=0$ is of the form
\vspace*{-4pt}\begin{align}\nonumber
o(1)&>\left[\min_{d:\lambda(1-\epsilon)\leq d\leq\lambda(1+\epsilon)}\left(\!1\!-\!\frac{m+1}{\log_2{n\choose d}}\!\right)\!\right]\!\!\sum_{d=\lambda(1-\epsilon)}^{\lambda(1+\epsilon)}\!\!P(D\!=\!d)\\\nonumber
&>c(n)\!\left(\!1\!-\!\frac{m+1}{\log_2{n\choose (1-\epsilon)\lambda}}\!\right)\left(1\!-\!\e^{-\lambda f_1(\epsilon)}\!-\!\e^{-\lambda f_2(\epsilon)}\right)\\\nonumber
&>c(n)\left(1-\frac{m+1}{\log_2{n\choose (1\!-\!\epsilon)\lambda}}-\e^{-\lambda f_1(\epsilon)}-\e^{-\lambda f_2(\epsilon)}\right).
\end{align}
\vspace*{1pt}

As a result, one has
\begin{equation}\nonumber
\left(\!1\!-\!\frac{m+1}{\log_2{n\choose (1-\epsilon)\lambda}}\!\right)\!<\!o(1)\Rightarrow m\!\geq\!\log_2\!{n\choose (1-\epsilon)\lambda}(1-o(1)).
\vspace*{1pt}
\end{equation}
This inequality can be further simplified as
\vspace*{1pt}\begin{align}\nonumber
m\!&\geq\!\log_2\!{n\choose (1-\epsilon)\lambda}(1-o(1))\\\nonumber
&\geq(1-\epsilon)\lambda(n)\log_2\frac{n}{(1-\epsilon)\lambda}\  (1-o(1))\\\nonumber
&=(1-\epsilon)\lambda(n)\log_2{n}\  (1-o(1)),
\vspace*{-4pt}\end{align}
where the last equality follows since $\lambda(n)=o(n)$.
\end{proof}

\vspace*{-9pt}
\subsection{Constructive upper Bounds on the minimum number of tests}

We describe next two nonadaptive methods for Poisson PGT and find the number of tests that ensures $\lim_{n\rightarrow\infty}P(\mathcal{E})=0$. For this purpose, we consider two separate asymptotic regimes for $\lambda(n)$: one, in which $\lambda(n)=o(n)$ and $\lim_{n\rightarrow\infty}\lambda(n)=\infty$, and another, in which $\lambda(n)=o(n)$ and $0<\lim_{n\rightarrow\infty}\lambda(n)<\infty$. Note that the case of constant $\lambda$ is covered by the latter scenario. 
We start by proving the following simple large deviations results, which we find useful in our subsequent derivations.
\begin{lemma}\label{lemma:2}
Let $D$ be a random variable following the right-truncated Poisson distribution, with $\lambda(n)=o(n)$ and $\lim_{n\rightarrow\infty}\lambda(n)=\infty$. Then, for any fixed $\epsilon>0$, one has $\lim_{n\rightarrow\infty}P(D>\Delta)=0$, where $\Delta=\lceil\lambda(n)^{(1+\epsilon)}\rceil-1$. 
\end{lemma}
\begin{proof}
Using Markov's inequality, one has
\vspace*{-4pt}
\begin{align}\nonumber
P(D\!>\!\Delta)\leq\frac{\mathbb{E}[D]}{\lceil\lambda(n)^{(1+\epsilon)}\rceil}&=\frac{\lambda(n)}{\lceil\lambda(n)^{(1+\epsilon)}\rceil}(1\!-\!o(1))\\\nonumber
&=\frac{1}{\lambda^{\epsilon}}(1+o(1))=o(1),
\end{align}
where the last claim follows since $\lim_{n\rightarrow\infty}\lambda(n)=\infty$. 
\end{proof}

Although this lemma is applicable when $\lim_{n\rightarrow\infty}\lambda(n)=\infty$, for the case when $0\!<\!\lim_{n\rightarrow\infty}\lambda(n)\!<\!\infty$ (including the case when $\lambda$ is a constant), the above arguments do not follow through. For this case, we prove a lemma in which a slowly-growing function of $n$, i.e. $\beta(n)\!=\!\log^{(K)}n$ defined in~\eqref{eq:logk}, is used to provide the needed guarantees. 

\begin{lemma}\label{lemma:2b}
Let $D$ be a random variable following the right-truncated Poisson distribution, with $\lambda(n)\!=\!o(n)$ and $0\!<\!\lim_{n\rightarrow\infty}\lambda(n)\!<\!\infty$. Also, let $\beta(n)\!=\!\log^{(K)}n$, for some finite $K\!>\!1$. Then, $\lim_{n\rightarrow\infty}P(D\!>\!\Delta)\!=\!0$ for $\Delta\!=\!\lceil\beta(n)\lambda(n)\rceil\!-\!1$. 
\end{lemma}
\begin{proof}
Before proving the lemma, 
Using Markov's inequality, one has
\vspace*{-4pt}
\begin{align}\nonumber
P(D\!>\!\Delta)\leq\frac{\mathbb{E}[D]}{\lceil\beta(n)\lambda(n)\rceil}&=\frac{\lambda(n)}{\lceil\beta(n)\lambda(n)\rceil}(1\!-\!o(1))\\\nonumber
&=\frac{1}{\beta(n)}(1+o(1))=o(1),
\end{align}
where the last equality follows since $\lim_{n\rightarrow\infty}\beta(n)=\infty$. 
\end{proof}

\vspace*{-4pt}
\subsubsection{Nonadaptive method I}\label{sec:method1}
In our first construction, we use disjunct codes to devise practical Poisson PGT schemes. We start with the following definition.

\begin{defin}[\textbf{Binary $\boldsymbol{\Delta}$-disjunct codes~\cite{KS64,DH06}}]\label{CGTdisjunct}
A binary $\Delta$-disjunct code for conventional GT is a code of length $m$ and size $n$, such that for any set of $\Delta+1$ codewords, $\mathcal{X}=\{\mathbf{x}_j\}_1^{\Delta+1}$, and for any codeword $\mathbf{x}_i\in\mathcal{X}$, there exists at least one coordinate $k$ such that $\mathbf{x}_i(k)=1$ and $\mathbf{x}_j(k)=0$, for some $\mathbf{x}_j\in\mathcal{X}$, where $j\neq i$.
\end{defin}

It is well known that binary $\Delta$-disjunct codes are capable of identifying up to $\Delta$ defectives in the conventional GT model. In addition, these codes are endowed with an efficient decoder with computational complexity $O(mn)$. The decoding procedure is based on the fact that a codeword corresponds to a defective if and only if its support is a subset of the support of the vector of test results, $\mathbf{y}$. Hence, given $\mathbf{y}$ and $\mathbf{C}$, the set of defectives may be identified with zero probability of error through
\vspace*{-4pt}
\begin{equation}\label{decoder:disjunct}
\hat{\mathcal{D}}=\{i:\text{supp}(\mathbf{x}_i)\subseteq\text{supp}(\mathbf{y})\},
\vspace*{-4pt}
\end{equation}
where $\mathbf{x}_i$ is the $i^{\text{th}}$ column of $\mathbf{C}$ and $\text{supp}(\cdot)$ stands for the support of a vector (i.e. the set of its nonzero entries). 

We consider a simple probabilistic construction for the test matrix: the entries of the test matrix follow an i.i.d. Bernoulli$(p)$ distribution, such that each entry of $\mathbf{C}$ is equal to $1$ with probability $p$, and $0$ with probability $1-p$. Let $\Delta=\Delta(n,\lambda(n))$ be a properly chosen function of $n$ and $\lambda(n)$. The idea is to identify $m$, $p$ and $\Delta$ so that $\mathbf{C}$ is a $\Delta-$disjunct matrix with high probability, while at the same time, the probability that the number of defectives exceeds $\Delta$ is small, as formally stated in the following theorem.

\begin{theorem}\label{thm3}
Assume that $D$ follows the right-truncated Poisson distribution, with $\lambda(n)=o(n)$ and $\lim_{n\rightarrow\infty}\lambda(n)=\infty$. Construct a test matrix by choosing each entry according to a Bernoulli$(p)$ distribution, where $p={(\lceil\lambda(n)^{(1+\epsilon)}\rceil)}^{-1}$, and where $\epsilon>0$ is arbitrarily small. Then $m= \e(\lceil\lambda(n)^{(1+\epsilon)}\rceil)^2\log n=\e\lambda(n)^{2(1+\epsilon)}\log n\: (1\!+\!o(1))$ tests suffice to ensure $\lim_{n\rightarrow\infty}P(\mathcal{E})=0$ using a decoding algorithm with computational complexity $O(mn)$. 
\end{theorem}
\begin{proof}
For any value of $\Delta>0$, we may write $P(\mathcal{E})$ as
\vspace*{-4pt}\begin{align}\nonumber
P(\mathcal{E})&=P(\mathcal{E}|D\leq\Delta)P(D\leq\Delta)+P(\mathcal{E}|D>\Delta)P(D>\Delta)\\\nonumber
&\leq P(\mathcal{E}|D\leq\Delta)+P(D>\Delta).
\vspace*{-4pt}\end{align}
From Lemma~\ref{lemma:2}, we know that $\Delta=\lceil\lambda(n)^{(1+\epsilon)}\rceil-1$ ensures $\lim_{n\rightarrow\infty}P(D>\Delta)=0$.
In order to bound $P(\mathcal{E}|D\leq\Delta)$, we use the following argument. The test matrix is constructed in a probabilistic, i.i.d. manner using the Bernoulli$(p)$ distribution. Given a fixed test matrix $\mathbf{C}$ and a vector of test results $\mathbf{y}$, we use the decoder in~\eqref{decoder:disjunct} to find $\hat{\mathcal{D}}$. Let $\mathcal{E}'$ be the event that $\mathbf{C}$ is not $\Delta$-disjunct. Since a $\Delta$-disjunct test matrix can identify up to $\Delta$ defectives with \emph{zero error probability}, then conditioned on $D\leq\Delta$, one has $\mathcal{E}\subseteq\mathcal{E}'$. As a result, $P(\mathcal{E}|D\leq\Delta)\leq P(\mathcal{E}'|D\leq\Delta)=P(\mathcal{E}')$, where the last equality follows since the events $\mathcal{E}'$ and $\{{D\leq\Delta\}}$ are independent.


It has been shown in~\cite[Thm. 8.1.3]{DH00} that by choosing $p=\frac{1}{\Delta+1}$ and $\pi_N=p{(1-p)}^\Delta$, one can bound $P(\mathcal{E}')$ as
\vspace*{-4pt}\begin{align}\nonumber
P(\mathcal{E}')&\leq (\Delta+1) {n\choose \Delta+1}{\left(1-\pi_N\right)}^m\\\nonumber
&\leq\exp(\!-{m\pi_N}\!+\!(\Delta\!+\!1)\!+\!(\Delta\!+\!1)\log n\!-\!\Delta\log(\Delta\!+\!1)).
\vspace*{-4pt}\end{align}
Hence, $\frac{(\Delta+1)}{\pi_N}\log n$ tests suffice to ensure $\lim_{n\rightarrow\infty} P(\mathcal{E}')=0$. Substituting $\pi_N=\frac{\Delta^{\Delta}}{{(\Delta+1)}^{\Delta+1}}$ yields
\vspace*{-4pt}\begin{align}\nonumber
\frac{(\Delta+1)}{\pi_N}\log n&=(\Delta+1)^{2}{\left(1+\frac{1}{\Delta}\right)}^{\Delta}\log n\\\nonumber
&\leq\e{(\Delta+1)}^2\log n=\e(\lceil\lambda(n)^{(1+\epsilon)}\rceil)^2\log n.
\vspace*{-4pt}\end{align}
In addition, since $P(\mathcal{E})\leq P(\mathcal{E}|D\leq\Delta)+P(D>\Delta)\leq  P(\mathcal{E}')+P(D>\Delta)$,  $m= \e(\lceil\lambda(n)^{(1+\epsilon)}\rceil)^2\log n=\e\lambda(n)^{2(1+\epsilon)}\log n\: (1\!+\!o(1))$ tests suffice to ensure $\lim_{n\rightarrow\infty} P(\mathcal{E})\!=\!0$. 
\end{proof}

The previous theorem relies on the assumption that $\lim_{n\rightarrow\infty}\lambda(n)\!=\!\infty$. A similar approach can be used for the case $0\!<\!\lim_{n\rightarrow\infty}\lambda(n)\!<\!\infty$, as described in the theorem to follow.
\begin{theorem}\label{thm4}
Assume that $D$ follows the right-truncated Poisson distribution, with $\lambda(n)=o(n)$ and $0<\lim_{n\rightarrow\infty}\lambda(n)<\infty$. Let $\beta(n)=\log^{(K)}n$, for some finite $K>1$. Construct a test matrix by choosing each entry according to a Bernoulli$(p)$ distribution, where $p={(\lceil\beta(n)\lambda(n)\rceil)}^{-1}$. Then $m= \e\;(\lceil\beta(n)\lambda(n)\rceil)^2\log n=\e\;(\beta(n)\lambda(n))^2\log n\:(1+o(1))$ tests suffice to ensure $\lim_{n\rightarrow\infty} P(\mathcal{E})=0$ using a decoding algorithm with computational complexity of $O(mn)$.
\end{theorem}
\begin{proof}
Similarly as in the proof of Theorem~\ref{thm3}, we may write $P(\mathcal{E})\leq P(\mathcal{E}|D\leq\Delta)+P(D>\Delta)$, for any $\Delta>0$. From Lemma~\ref{lemma:2b}, we know that setting $\Delta=\lceil\beta(n)\lambda(n)\rceil-1$ ensures $\lim_{n\rightarrow\infty}P(D>\Delta)=0$. By choosing $p=\frac{1}{\Delta+1}$ and invoking the same arguments  as those in Theorem~\ref{thm3}, we conclude that $m=\e\;(\Delta+1)^2\log n$ tests suffice for $\lim_{n\rightarrow\infty}P(\mathcal{E}|D\leq\Delta)=0$. Substituting the previously computed value of $\Delta$ into the expression for the number of tests results in $m=\e\;(\lceil\beta(n)\lambda(n)\rceil)^2\log n$.
\end{proof}

Theorems~\ref{thm3} and~\ref{thm4} do not account for the presence of errors in the vector of test results. In order to address this issue, we invoke the following definition of an error-tolerent binary disjunct code.

\begin{defin}[\textbf{Error tolerant binary $\boldsymbol{\Delta}$-disjunct codes~\cite{DH06}}]\label{CGTdisjunct2}
A binary $\Delta$-disjunct code designed for conventional GT, capable of correcting up to $v$ errors, is a code of length $m$ and size $n$ such that for any set of $\Delta+1$ codewords, $\mathcal{X}=\{\mathbf{x}_j\}_1^{\Delta+1}$, and for any codeword $\mathbf{x}_i\in\mathcal{X}$, there exists 
a set of coordinates $\mathcal{R}_i$ of size at least $2v+1$, such that $\forall k\in\mathcal{R}_i$, $\mathbf{x}_i(k)=1$ and $\mathbf{x}_j(k)=0$, for some $\mathbf{x}_j\in\mathcal{X}$ with $j\neq i$.
\end{defin}

In order to identify the set of defectives using these codes with a zero error probability, we use the following decoder. For each codeword $\mathbf{x}_i$, $i\in\{1,2,\dots,n\}$, let $N_i$ denote the number of coordinates $j\in\{1,2,\dots,m\}$ for which $\mathbf{x}_i(j)=1$ and $\mathbf{y}(j)=0$ hold simultaneously. Then
\vspace*{-4pt}
\begin{equation}\label{decoder:disjunct2}
\hat{\mathcal{D}}=\{i:N_i\leq v\}.
\vspace*{-4pt}
\end{equation}
Note that the computational complexity of this decoding method is $O(mn)$. 
The next theorems use error-tolerant  disjunct codes in order to bound the number of tests for a Poisson PGT model that guarantees~\eqref{condition1} in the presence of up to $v$ errors in the vector of test results $\mathbf{y}$.

\begin{theorem}\label{thm5}
Assume that the number of defectives follows the right-truncated Poisson distribution, with $\lambda(n)=o(n)$ and $\lim_{n\rightarrow\infty}\lambda(n)=\infty$. Construct a test matrix by choosing each entry according to a Bernoulli$(p)$ distribution, where $p={(\lceil\lambda(n)^{(1+\epsilon)}\rceil)}^{-1}$ and $\epsilon>0$ is arbitrarily small. Then $m= \left(2\e\lambda(n)^{2(1+\epsilon)}\log {n}+4\e v(n)\lambda(n)^{1+\epsilon}\right)(1+o(1))$ tests suffice to ensure $\lim_{n\rightarrow\infty}P(\mathcal{E})=0$ in the presence of not more than $v(n)$ errors, using a decoding algorithm with computational complexity $O(mn)$. 
\end{theorem}

\begin{proof}
Similar to the proof of Theorem~\ref{thm3}, we may write $P(\mathcal{E})\leq P(\mathcal{E}|D\leq\Delta)+P(D>\Delta)$, for any value of $\Delta>0$. Lemma~\ref{lemma:2} can be used directly to show that $\lim_{n\rightarrow\infty}P(D>\Delta)=0$, if $\Delta=\lceil\lambda(n)^{(1+\epsilon)}\rceil-1$. In order to bound $P(\mathcal{E})\leq P(\mathcal{E}|D\leq\Delta)$, the approach of~\cite[Thm. 8.1.3]{DH00} used in Theorem~\ref{thm3} can be generalized to show that $P(\mathcal{E}|D\leq\Delta)\leq P(\mathcal{E}'|D\leq\Delta)=P(\mathcal{E}')$, where $\mathcal{E}'$ is the event that $\mathbf{C}$ is not an $v$ error correcting $\Delta$-disjunct test matrix. To bound $P(\mathcal{E}')$, we first fix a set of column-indices $\mathcal{I}:|\mathcal{I}|=\Delta+1$ and let $k\in\mathcal{I}$ be fixed. There are $(\Delta+1) {n\choose \Delta+1}$ ways to choose $k$ and $\mathcal{I}$. For a fixed choice of $\mathcal{I}$ and $k$, $\forall j\in \{1,2,\dots,m\}$, let $N_j$ be a Bernoulli random variable such that it has a value $1$ if the $j^{\text{th}}$ row of $\mathbf{C}$ has a value $1$ in the $k^{\text{th}}$ column while having $0$ in each column indexed by $\mathcal{I}\backslash\{k\}$, and $N_j$ has a value $0$ otherwise. By definition, the random variables $N_j$ are i.i.d., and for $j\in m$ one has
\vspace*{-4pt}\begin{equation}\nonumber
\Pro(N_j=1)={p{(1-p)}^\Delta}\triangleq\pi_N. 
\vspace*{-4pt}\end{equation}
Using the Chernoff bound for Binomial random variables for $0<\delta<1$, one obtains
\vspace*{-4pt}\begin{equation}\nonumber
\Pro\left(\sum_{j=1}^mN_j\leq(1-\delta)m\pi_N\right)\leq\exp\left(-\frac{\delta^2m\pi_N}{2}\right).
\vspace*{-4pt}\end{equation}
By setting $\delta=1-\frac{2v}{m\pi_N}$, it follows that
\vspace*{-4pt}\begin{equation}\nonumber
\Pro\left(\sum_{j=1}^mN_j\leq2v\right)\leq\exp\left({-\frac{m\pi_N}{2}{\left(1-\frac{2v}{m\pi_N}\right)^2}}\right),
\vspace*{-4pt}\end{equation}
which provides an upper bound on the probability that for a fixed $\mathcal{I}$ and $k$, at most $2v$ rows of $\mathbf{C}$ satisfy the disjunctness property. As a result,
\vspace*{-4pt}\begin{align}\nonumber
&P(\mathcal{E}')\leq{n\choose \Delta+1}(\Delta+1)\exp\left({-\frac{m\pi_N}{2}{\left(1-\frac{2v}{m\pi_N}\right)^2}}\right)\\\nonumber
&\leq\!\exp(\!(\!\Delta\!+\!1\!)\!\log n\!+\!\Delta\!+\!1\!-\!\Delta\!\log(\Delta\!+\!1)\!-\!\frac{m\pi_{\!N}}{2}\!-\!\frac{2v^2}{m\pi_{\!N}}\!+\!2v\!).
\vspace*{-4pt}\end{align}
Hence, $2\frac{(\Delta+1)}{\pi_N}\log n+\frac{4v}{\pi_N}$ tests suffice to ensure $\lim_{n\rightarrow\infty} P(\mathcal{E}')=0$. Substituting $\pi_N=\frac{\Delta^{\Delta}}{{(\Delta+1)}^{\Delta+1}}$, yields
\vspace*{-4pt}\begin{align}\nonumber
2\frac{(\Delta\!+\!1)}{\pi_N}\log n\!&+\!\frac{4v}{\pi_N}\!=\!2(\Delta\!+\!1){\left(\!1\!+\!\frac{1}{\Delta}\!\right)}^{\Delta}\left((\Delta\!+\!1)\log n\!+\!{2v}\right)\\\nonumber
&\leq \!2\e(\Delta\!+\!1)\left((\Delta\!+\!1)\log n\!+\!{2v}\right)\\\nonumber
&=\!2\e\;(\lceil\lambda(n)^{(1+\epsilon)}\rceil)\left((\lceil\lambda(n)^{(1+\epsilon)}\rceil)\log n\!+\!{2v}\right).
\vspace*{-4pt}\end{align}
Consequently, $m\!=\!\left(2\e\lambda(n)^{2(1\!+\!\epsilon)}\log {n}\!+\!4\e v(n)\lambda(n)^{1\!+\!\epsilon}\right)(1\!+\!o(1))$,
tests suffice to ensure $\lim_{n\rightarrow\infty} P(\mathcal{E})\!=\!0$. 
\end{proof}

\begin{theorem}\label{thm5b}
Assume that the number of defectives follows the right-truncated Poisson distribution, with $\lambda(n)=o(n)$ and $0<\lim_{n\rightarrow\infty}\lambda(n)<\infty$. Let $\beta(n)=\log^{(K)}n$, for some finite $K>1$. Construct a test matrix by choosing each entry according to a Bernoulli$(p)$ distribution, where $p={(\lceil\beta(n)\lambda(n)\rceil)}^{-1}$. Then $m=\left({2\e\; (\beta(n)\lambda(n))^2}\log n+4\e v(n)\beta(n)\lambda(n)\right)(1+o(1))$ tests suffice to ensure $\lim_{n\rightarrow\infty}P(\mathcal{E})=0$ in the presence of not more than $v(n)$ errors, using a decoding algorithm with computational complexity $O(mn)$. 
\end{theorem}
\begin{proof}
Since $P(\mathcal{E})\leq P(\mathcal{E}|D\leq\Delta)+P(D>\Delta)$, for any $\Delta>0$, Lemma~\ref{lemma:2b} ensures that $\lim_{n\rightarrow\infty}P(D>\Delta)=0$ if $\Delta=\lceil\beta(n)\lambda(n)\rceil-1$. Repeating the arguments of Theorem~\ref{thm5}, we conclude that $m=2\e{(\Delta+1)}^2\log n+{4\e v}{(\Delta+1)}$ tests suffice for $\lim_{n\rightarrow\infty}P(\mathcal{E}|D\leq\Delta)=0$. Substituting the previously computed value of $\Delta$ into the expression for the number of tests results in $m=\left({2\e(\beta(n)\lambda(n))^2}\log n+4\e v(n)\beta(n)\lambda(n)\right)(1+o(1))$.
\end{proof}

\vspace*{-4pt}
\subsubsection{Nonadaptive method II}\label{sec:method2}
In~\cite{CD08}, Cheng and Du described the construction of a probabilistic test matrix for the nonadaptive combinatorial GT model, and proved that their test matrix can identify up to $\Delta$ defectives from $n$ subjects with high probability. Although the underlying codes are not binary disjunct codes, the decoder in \eqref{decoder:disjunct} can be used to identify the defectives with high probability. The construction consists of two steps: in the first step, a nonbinary test matrix with i.i.d. entries is created; in the second step, a transformation is used to convert this nonbinary matrix into a binary matrix~\cite[Thm. 1]{CD08}. One should note that as a consequence of this transformation, the entries of the binary test matrix are no longer i.i.d. We use this construction technique to identify the set of defectives in Poisson PGT, and achieve this with a suitable choice of $\Delta$. The following lemma is a restatement of the results in~\cite[Thm. 10]{CD08}, suitable for our application.

\begin{lemma}\label{lemma:CD08}
The nonadaptive group testing method in~\cite{CD08} can identify up to $\Delta$ defectives among $n$ subjects, using no more than $\frac{3\Delta}{\log_2 3}\!\left(\!\log_2 \!n\!+\!\log_2\!\frac{1}{1-p}\!\right)$ tests, with probability at least $p$. 
\end{lemma}
\begin{proof}
See~\cite[Thm. 10]{CD08} and its proof. 
\end{proof}
Next, we show how this pooling design can be used to identify the set of defectives in Poisson PGT, while ensuring a probability of error that diminishes asymptotically.

\begin{theorem}\label{thm_best}
Assume that $D$ follows the right-truncated Poisson distribution, with $\lambda(n)=o(n)$ and $\lim_{n\rightarrow\infty}\lambda(n)=\infty$. Then, one can identify the set of defectives such that $\lim_{n\rightarrow\infty}P(\mathcal{E})=0$, using $m= \frac{3}{\log_2 3}\lambda(n)^{(1+\epsilon)}\log_2 n\:(1+o(1))$ tests.\end{theorem}
\begin{proof}
We first write $P(\mathcal{E})$ as
\vspace*{-4pt}\begin{align}\nonumber
P(\mathcal{E})&=P(\mathcal{E}|D\leq\Delta)P(D\leq\Delta)+P(\mathcal{E}|D>\Delta)P(D>\Delta)\\\nonumber
&\leq P(\mathcal{E}|D\leq\Delta)+P(D>\Delta).
\vspace*{-4pt}\end{align}
Given $\Delta=\lceil\lambda(n)^{(1+\epsilon)}\rceil-1,$ for a fixed $\epsilon>0$, we use Lemma~\ref{lemma:2} to conclude that $\lim_{n\rightarrow\infty}P(D>\Delta)=0$. By setting $p=1-\frac{1}{\log n}$ and using Lemma~\ref{lemma:CD08}, we can show that one can identify up to $\Delta$ defectives with no more than $m=\frac{3\Delta}{\log_2 3}\log_2 n\left(1+o(1)\right)$ tests, so that the probability of error is bounded as
\vspace*{-4pt}\begin{equation}\nonumber
P(\mathcal{E}|D\leq\Delta)\leq 1-p=\frac{1}{\log n}.
\vspace*{-4pt}\end{equation}
Consequently, one has
\vspace*{-4pt}\begin{equation}\nonumber
\lim_{n\rightarrow\infty}P(\mathcal{E})\leq \lim_{n\rightarrow\infty}P(\mathcal{E}|D\leq\Delta)+\lim_{n\rightarrow\infty} P(D>\Delta)=0.
\vspace*{-4pt}\end{equation}
\end{proof}

\begin{theorem}\label{thm_best2}
Assume that $D$ follows the right-truncated Poisson distribution, with $\lambda(n)=o(n)$ and $0<\lim_{n\rightarrow\infty}\lambda(n)<\infty$. Let $\beta(n)=\log^{(K)}n$, for some value of $K>1$. Then, one can identify defectives with $\lim_{n\rightarrow\infty}P(\mathcal{E})=0$, using $m= \frac{3}{\log_2 3}\,\beta(n)\lambda(n)\log_2 n\:(1+o(1))$ tests.
\end{theorem}
\begin{proof}
Similar to what was done in Theorem~\ref{thm_best}, we may write $P(\mathcal{E})\leq P(\mathcal{E}|D\leq\Delta)+P(D>\Delta)$, for any value of $\Delta>0$. Given $\Delta=\lceil\beta(n)\lambda(n)\rceil-1$, we use Lemma~\ref{lemma:2b} to conclude that $\lim_{n\rightarrow\infty}P(D>\Delta)=0$. By setting $p=\frac{1}{\Delta+1}$ and invoking the same arguments provided in the proof of Theorem~\ref{thm_best}, we conclude that $m=\frac{3\Delta}{\log_2 3}\,\log_2 n\left(1+o(1)\right)$ tests are sufficient to ensure that $\lim_{n\rightarrow\infty}P(\mathcal{E}|D\leq\Delta)=0$. Substituting the previously computed value of $\Delta$ into the expression for $m$ results in $m= \frac{3}{\log_2 3}\,\beta(n)\lambda(n)\log_2 n\:(1+o(1))$.
\end{proof}

\vspace*{-5pt}
\section{Semi-adaptive methods for Poisson PGT}\label{sec:adaptive}

An alternative to both adaptive and non-adaptive GT approaches is \emph{semi-adaptive} testing. A semi-adaptive GT algorithm is an algorithm in which tests are designed in several stages. The tests in each stage are constructed in a nonadaptive manner and therefore can be performed in parallel. However, the set of subjects on which the tests are preformed changes from one stage to the next; in other words, the results obtained during one stage of testing may guide the choice of test subjects and potential defectives in the next stage. One of the best known semi-adaptive algorithms is the original $2$-stage algorithm proposed by Dorfman~\cite{D43}. 

In the absence of error, a semi-adaptive algorithm is expected to identify all defectives, even if no prior knowledge regarding the number of defectives is available. As a result, unlike the case of nonadaptive algorithms in which one seeks to find a number of tests $m$ for which $\lim_{n\rightarrow\infty}P(\mathcal{E})=0$, in semi-adaptive framework one is interested in the expected number of tests $\bar{m}$ that an algorithm performs in order to identify the defectives with zero probability of error, i.e., with $P(\mathcal{E})=0$. In what follows, we first find a lower bound on $\bar{m}$ for any adaptive (and hence, semi-adaptive) algorithm for Poisson PGT using Huffman coding. Then, we devise a semi-adaptive algorithm and show that for this algorithm, $\bar{m}$ is only a constant factor away from the lower bound. 

\vspace*{-5pt}
\subsection{Lower bound on the expected number of tests}\label{sec:lower_bound_adaptive}
Suppose that the number of defectives follows the truncated Poisson distribution; in addition, assume that for any fixed $1\leq d\leq n$, all the sets of $D=d$ defectives are equally likely. 

In what follows, we show that one can use Huffman source coding~\cite{CT91} to find a lower bound on the expected number of adaptive tests required to identify the defectives. Let $\mathbf{w}\in\{0,1\}^n$ be a binary random vector such that $\mathbf{w}(i)=1$ if the $i^{\text{th}}$ subject is a defective, and $\mathbf{w}(i)=0$ otherwise. There are $2^n$ choices for $\mathbf{w}$, contained in a set denoted by $\mathcal{W}$. An adaptive GT algorithm has to identify the true realization of $\mathbf{w}$, denoted by $\mathbf{w}_t$, using a set of tests. Each such test can be represented as a ``yes/no'' query of the form ``is $\mathbf{w}_t$ a member of the set $\mathcal{W}'$?'', where the set $\mathcal{W}'\subseteq\mathcal{W}$ is determined by the design of the test. For example for $n=5$, the query corresponding to a test that contains the first, the fourth and the fifth subjects asks if $\mathbf{w}_t\in\mathcal{W}'$, where
\vspace*{-4pt}\begin{equation}\nonumber
\mathcal{W}'=\{{\begin{bmatrix} 0\\0\\0\\0\\0
\end{bmatrix},\begin{bmatrix} 0\\1\\0\\0\\0
\end{bmatrix},\begin{bmatrix} 0\\0\\1\\0\\0
\end{bmatrix},\begin{bmatrix} 0\\1\\1\\0\\0
\end{bmatrix}\}}.
\vspace*{-4pt}\end{equation}
If the output of the test is $0$, the answer to the query is ``yes'', since none of the three subjects in the test are defective and therefore $\mathbf{w}_t\in\mathcal{W}'$; otherwise the answer to the query is ``no'' which implies that $\mathbf{w}_t\in\mathcal{W}\backslash\mathcal{W}'$.
On the other hand, it can be easily verified that not every possible subset query corresponds to a group test~\cite{DH00,A93}. As a result, the minimum expected number of subset queries, required to identify $\mathbf{w}_t$, provides a lower bound on the minimum expected number of group tests required to identify $\mathbf{w}_t$ in an adaptive manner. One should note that the minimum expected number of queries of the form above is equal to the expected length of a Huffman code designed for a source with alphabet $\mathcal{W}$ and the corresponding probability distribution~\cite{CT91}.

For a fixed $0\leq d\leq n$, let $\mathbf{w}_{d,j}$, $j=1,2,\dots,{n\choose d}$, be a realization of $\mathbf{w}$ with exactly $d$ entries equal to $1$. As a result, the alphabet of the source $\mathbf{w}$ is of the form $\mathcal{W}=\{\mathbf{w}_{d,j}\}$, $j=1,2,\dots,{n\choose d}$, $d=0,1,\dots,n$. It follows that for all $0\leq d\leq n$ and for all $1\leq j\leq {n\choose d}$, 
\vspace*{-4pt}\begin{align}\nonumber
P(\mathbf{w}=\mathbf{w}_{d,j})&=\frac{1}{{n\choose d}} P(D=d)\\\label{vecdist}
&=c(n)\frac{(n-d)!}{n!}{{\lambda(n)}^d}\e^{-\lambda(n)}\triangleq P(\mathbf{w}_{d}).
\end{align}
\vspace{-0.3in}

\begin{theorem}\label{theorem:lower_exp}
Let $\lambda(n)=o(n)$. Then, the minimum expected number of tests in an adaptive (and semi-adaptive) group testing algorithm satisfies $\bar{m}>\lambda(n)\log_2\!\frac{n}{\lambda(n)}\  (1+o(1))-\log_2\!\e\  \frac{\lambda(n)^4}{n^2}$. In addition, if $\lambda(n)=o\left((n^2\log_2n)^{1/3}\right)$, this lower bound simplifies to $\bar{m}>\lambda(n)\log_2\frac{n}{\lambda(n)}\  (1+o(1))$.
\end{theorem}
\vspace{-0.2in}
\begin{proof}
To prove this theorem, we note that the Shannon entropy \cite{CT91} of the source, $H(\mathbf{w})$, provides a lower bound on the average length of the optimum Huffman code. Consequently, using~\eqref{vecdist}, one has
\vspace*{-4pt}\begin{align}\nonumber
\bar{m}\geq H(\mathbf{w})&= -\sum_{d=0}^n\sum_{j=1}^{{n\choose d}} P(\mathbf{w}_{d,j})\log_2P(\mathbf{w}_{d,j})\\\nonumber
&=-\sum_{d=0}^n{n\choose d} P(\mathbf{w}_{d})\log_2P(\mathbf{w}_{d})\\\nonumber
&=-\sum_{d=0}^nP(D=d)\log_2P(\mathbf{w}_{d})\\\nonumber
&=\mathbb{E}_D\left[ \log_2\frac{1} {P(\mathbf{w}_{d})}\right].
\vspace*{-4pt}\end{align}
By invoking~\eqref{vecdist}, the previous expression may be rewritten as
\vspace*{-4pt}\begin{align}\nonumber
\bar{m}&\geq\mathbb{E}_D\!\left[\lambda(n)\log_2\e\!-\!\log_2c(n)\!+\!\log_2\!\frac{n!}{(n-d)!}\!-\!d\log_2\lambda(n)\right]\\\nonumber
&\geq \lambda(n)\log_2\e\!-\!\log_2c\!+\!\mathbb{E}_D\!\left[d\log_2(n\!-\!d\!+\!1)\!-\!d\log_2\lambda(n)\right]\\\nonumber
&= \lambda\log_2\e\!-\!\log_2c\!+\!\log_2\!\frac{n}{\lambda}\mathbb{E}[D]\!+\!\mathbb{E}_D\!\!\left[\!d\log_2\!\left(\!1\!-\!\frac{d\!-\!1}{n}\!\right)\!\right]\\\label{eq:lower1}
&\geq \lambda(n)\log_2\e-\log_2c+\log_2\frac{n}{\lambda(n)}\  \mathbb{E}[D]\\\nonumber
&\  \  \  \  \  \  \  \  \  \  \  \  \  \  \  \  \  \  \  \  \  \  \  \  \  \  \  \  \  \  \  \  \  \  \  \  \  -\log_2\e\:\mathbb{E}_D\left[\frac{d(d-1)}{n-d+1}\right],
\vspace*{-4pt}\end{align}
where the last inequality follows since $\log(1+x)\geq\frac{x}{1+x}$, for any $x>-1$.
Next, note that 
\vspace*{-4pt}\begin{align}\nonumber
\mathbb{E}_D\left[\frac{d(d-1)}{n-d+1}\right]&=c(n)\sum_{d=0}^n\frac{d(d-1)}{n-d+1}\frac{\lambda(n)^d}{d!}\e^{-\lambda(n)}\\\label{eq:mean1}
&=\lambda(n)^2c(n)\sum_{d=0}^{n-2}\frac{\lambda(n)^d}{d!(n-d-1)}\e^{-\lambda(n)}.
\vspace*{-4pt}\end{align}
For any $d$ such that $0\leq d\leq n-2$, one has
\vspace*{-4pt}\begin{align}\nonumber
\frac{1}{n-d-1}&=\frac{1}{n}\sum_{i=0}^\infty{\left(\frac{d+1}{n}\right)}^i\\\nonumber
\end{align}

\begin{align}\nonumber
&=\frac{1}{n}\left(1+\frac{d+1}{n}+\frac{(d+1)^2}{n^2}\sum_{i=0}^\infty{\left(\frac{d+1}{n}\right)}^i\right)\\\nonumber
&=\frac{1}{n}\left(1+\frac{d+1}{n}+\frac{(d+1)^2}{n(n-d-1)}\right)\\\nonumber
&\leq\frac{1}{n}\left(1+\frac{d+1}{n}+\frac{(d+1)^2}{n}\right)\\\label{eq23}
&=\frac{1}{n^2}\left((n+2)+4d+d(d-1)\right).
\vspace*{-4pt}\end{align}
Substituting~\eqref{eq23} in~\eqref{eq:mean1} yields
\vspace*{-4pt}\begin{align}\nonumber
&\mathbb{E}_D\left[\frac{d(d-1)}{n-d+1}\right]= c\lambda(n)^2\frac{n+2}{n^2}\sum_{d=0}^{n-2}\frac{\lambda(n)^d}{d!}\e^{-\lambda(n)}\\\nonumber
&\  \  \  +4c\frac{\lambda(n)^3}{n^2}\sum_{d=0}^{n-3}\frac{\lambda(n)^d}{d!}\e^{-\lambda(n)}+c\frac{\lambda(n)^4}{n^2}\sum_{d=0}^{n-4}\frac{\lambda(n)^d}{d!}\e^{-\lambda(n)}\\\nonumber
&\  \  \  <\frac{\lambda(n)^2}{n^2}(n+2+4\lambda(n)+\lambda(n)^2)\\\label{eq:mean2}
&\  \  \  =\frac{\lambda(n)^2}{n}(1+o(1))+\frac{\lambda(n)^4}{n^2}.
\vspace*{-4pt}\end{align}
Substituting~\eqref{eq:mean0} and~\eqref{eq:mean2} in~\eqref{eq:lower1}, and by recalling that $\lambda(n)=o(n)$, one arrives at
\vspace*{-4pt}\begin{align}\nonumber
\bar{m}&\!\geq\! \lambda(n)\log_2\!\e\!-\!\log_2c\!+\!\log_2\!\frac{n}{\lambda(n)} \mathbb{E}[D]\!-\!\log_2\!\e\mathbb{E}_D\!\!\left[\!\frac{d(d\!-\!1)}{n\!-\!d\!+\!1}\!\right]\\\nonumber
&>\lambda(n)\log_2\!\frac{n}{\lambda(n)}\  (1+o(1))-\log_2\e\:\frac{\lambda(n)^4}{n^2}.
\vspace*{-4pt}\end{align}
If $\lambda(n)=o\left((n^2\log_2n)^{1/3}\right)$, this bound simplifies to $\bar{m}>\lambda(n)\log_2\frac{n}{\lambda(n)}\  (1+o(1))$.
\end{proof}

\vspace*{-5pt}
\subsection{Constructive upper bound on expected number of tests using an $s$-stage algorithm}
In~\cite{L62}, Li proposed an $s$-stage algorithm to identify $d$ defectives in a combinatorial group testing framework. In what follows, we modify his algorithm and show that the expected number of tests performed by $s$-stage testing allows one to find all the defectives in a Poisson PGT model, while being only a constant away from the lower bound of Theorem~\ref{theorem:lower_exp}. 

Let $s=s(n,\lambda(n))$ denote the total number of stages. Also, let $\mathcal{S}_i$, $1\leq i\leq s$, be the set of potential defectives at stage $i$ on which the group tests are performed. In the first stage, we set $\mathcal{S}_1=\mathcal{S}$, where $\mathcal{S}$ is the set of all subjects, $|\mathcal{S}|=n$. Then, we randomly divide $\mathcal{S}_1$ into disjoint sets of size $k_1$, where $k_1=k_1(\lambda(n),n)$. If $k_1$ does not divide $|\mathcal{S}_1|$, one set will contain fewer than $k_1$ entries, equal to the remainder of dividing $|\mathcal{S}_1|$ by $k_1$. A test is performed on each of these sets independently. In the second stage, $\mathcal{S}_2$ is formed by pooling all the subjects in sets with a positive test outcome in the first stage. Similarly, the set $\mathcal{S}_2$ is randomly divided into disjoint sets of size $k_2$. Again, one set may contain fewer subjects as compared to the other sets, and a test is performed on each set. The procedure continues in the same manner up to stage $s\!-\!1$. In the last stage, $\mathcal{S}_s$ is formed by pooling all the subjects in sets with a positive test outcome at stage $s\!-\!1$; then, each remaining subject is tested individually to determine if it is defective. The following theorem shows that proper choices of $s$ and $k_i$, $1\!\leq \!i\!\leq \!s\!-\!1$, may guarantee that the expected number of tests performed using this algorithm is upper bounded by a value only a constant away from the lower bound. 

\begin{theorem}\label{theorem:upper_exp}
Let $\lambda(n)=o(n)$ and let $\bar{\lambda}(n)=\mathbb{E}_{D}[d]$. Then, by choosing $s_0=\log\frac{n}{\bar{\lambda}(n)}$, $s=\left\lceil s_0\right\rceil$, and $k_i=\left\lceil{\left(\frac{n}{\bar{\lambda}(n)}\right)}^{\frac{s_0-i}{s_0}}\right\rceil$, for $1\leq i\leq s-1$, the expected number of the proposed semi-adaptive group testing algorithm satisfies 
\begin{align}\nonumber
\bar{m}&\leq\!\frac{\e}{\log_2\e}\bar{\lambda}(n)\log_2\!\left(\frac{n}{\bar{\lambda}(n)}\right)\:(1\!+\!o(1))\\\nonumber
&=\frac{\e}{\log_2\e}{\lambda(n)}\log_2\!\left(\frac{n}{{\lambda(n)}}\right)\:(1\!+\!o(1)),
\end{align}
where $\frac{\e}{\log_2\e}\approx 1.884$.
\end{theorem}
\begin{proof}
Assume that $D=d$ is the number of defectives. In the first stage, divide the test subjects into disjoint groups of size $k_1$. This leads to $\left\lceil\frac{n}{k_1}\right\rceil$ tests. In the $i^{\text{th}}$ stage, $1\leq i\leq s-2$, at most $d$ tests are positive, with the upper bound achieved when each defective is in a different group; as a result, the number of remaining subjects and the number of tests in the $(i+1)^{\text{th}}$ stage are at most $dk_i$ and $\left\lceil d\frac{k_{i}}{k_{i+1}}\right\rceil$, respectively. In the last stage, the number of remaining subjects and the number of tests both equal to $dk_{s-1}$. Hence, the total number of tests is bounded as
\vspace*{-4pt}\begin{equation}\nonumber
m\leq \left\lceil\frac{n}{k_1}\right\rceil+\sum_{i=2}^{s-1}\left\lceil d\frac{k_{i-1}}{k_i}\right\rceil+d\:k_{s-1}.
\vspace*{-4pt}\end{equation}
Consequently since $s$ and $k_i$, $1\leq i\leq s-1$, do not depend on $d$, one has
\vspace*{-4pt}\begin{align}\nonumber
\bar{m}=\mathbb{E}_{D}[m]&\leq\left\lceil\frac{n}{k_1}\right\rceil+\sum_{i=2}^{s-1}\mathbb{E}_D\left[\left\lceil d\frac{k_{i-1}}{k_i}\right\rceil\right]+\bar{\lambda}(n)\:k_{s-1}\\\nonumber
&\leq\left\lceil\frac{n}{k_1}\right\rceil+\sum_{i=2}^{s-1}\mathbb{E}_D\left[d\frac{k_{i-1}}{k_i}+1\right]+\bar{\lambda}(n)\:k_{s-1}\\\nonumber
&=\left\lceil\frac{n}{k_1}\right\rceil+\sum_{i=2}^{s-1}\bar{\lambda}(n)\frac{k_{i-1}}{k_i}+\bar{\lambda}(n)\:k_{s-1}+s-2.
\vspace*{-4pt}\end{align}
Substituting $s$ and $k_i$ in the previous expressions, one obtains
\vspace*{-4pt}\begin{align}\nonumber
\bar{m}&\leq\left\lceil\frac{n}{k_1}\right\rceil+\sum_{i=2}^{s-1}\bar{\lambda}(n)\frac{k_{i-1}}{k_i}+\bar{\lambda}(n)\:k_{s-1}+s-2\\\nonumber
&\leq\bar{\lambda}(n){\left(\!\frac{n}{\bar{\lambda}(n)}\!\right)}^{\!\frac{1}{s_0}}\!\:(1\!+\!o(1))\!+\!\sum_{i=2}^{s-1}\bar{\lambda}(n)\!\!\: {\left(\!\frac{n}{\bar{\lambda}(n)}\!\right)}^{\!\frac{1}{s_0}}(1\!+\!o(1))\\\nonumber
&\  \  \  \  \  \  \  \  \  \  \  \  \  \  \  \  \  \  \  \  \  \  \  \  +\!\bar{\lambda}(n)\!\:{\left(\!\frac{n}{\bar{\lambda}(n)}\!\right)}^{\!\frac{2}{s_0}}\!+\!\log\frac{n}{\bar{\lambda}(n)}(1\!+\!o(1))\\\nonumber
&\leq\e(s-1)\bar{\lambda}(n)\!\:(1\!+\!o(1))+\bar{\lambda}(n)\!\:\e^2+\log\frac{n}{\bar{\lambda}(n)}(1\!+\!o(1))\\\nonumber
&\leq\e\bar{\lambda}(n)\!\:\log\frac{n}{\bar{\lambda}(n)}\!\:(1+o(1))\\\nonumber
&=\frac{\e}{\log_2\e}\bar{\lambda}(n)\log_2\left(\!\frac{n}{\bar{\lambda}(n)}\!\right)\!\:(1+o(1)).
\vspace*{-4pt}\end{align}
\end{proof}

\vspace*{-5pt}
\section{Summary of the results and discussion}\label{discussion}
In the previous sections, we introduced the Poisson probabilistic group testing framework for modeling the number of defectives according to a random variable following a right-truncated Poisson distribution. 
For the proposed model and under the assumption that $\lambda(n)=o(n)$, we considered nonadaptive and semi-adaptive methods to identify the defectives. These methods are based on generalization of combinatorial GT schemes, which to the best of our knowledge are used in the context of probabilistic GT for the first time. 

In Section~\ref{sec:lower_bound}, we used information theoretic arguments to derive a lower bound on the number of tests (Thm.~\ref{theorem:lower}). In addition, we derived constructive upper bounds on the number of tests using practical testing schemes (Thms.~\ref{thm3}-\ref{thm_best2}) and information theoretic arguments (Thms.~\ref{thm1} and \ref{thm2}). It is worth mentioning that if $\lambda(n)$ grows slowly with $n$ (i.e. $\log^3(n)=o(\log n/\beta(n)))$, a conclusion of Thm.~\ref{thm1} is that $2 \lambda(n)^{1+\alpha}\log n$ measurements is sufficient to find the defectives when $\lim_{n\rightarrow\infty}\lambda(n)=\infty$. Similar simplifications can also be obtained for the case where $0<\lim_{n\rightarrow\infty}\lambda(n)<\infty$ using Thm.~\ref{thm2}. The results under the assumption that the vector of test results is error-free are summarized in Table~\ref{table:nonadaptive}. In the table, $\beta(n)$ is used to represent the slowly-growing function defined in~\eqref{eq:logk}, and $\epsilon$, $\alpha$, $\delta$, and $\gamma$ are arbitrary small positive constants. In Thms.~\ref{thm5} and~\ref{thm5b}, we considered the case in which there are at most $v(n)$ errors in the vector of test results and showed that $m=\left({2\e\; (\beta(n)\lambda(n))^2}\log n+4\e v(n)\beta(n)\lambda(n)\right)(1+o(1))$ tests are sufficient to identify the defectives using a decoder with computational complexity of $O(mn)$ if $0<\lim_{n\rightarrow\infty}\lambda(n)<\infty$. Similarly, we showed that if $\lim_{n\rightarrow\infty}\lambda(n)=\infty$, the same decoder requires $m= \left(2\e\lambda(n)^{2(1+\epsilon)}\log {n}+4\e v(n)\lambda(n)^{1+\epsilon}\right)(1+o(1))$ tests.

The test constructions and decoding algorithms used in Thms.~\ref{thm3}-\ref{thm5b} rely on designing test matrices that can identify the defectives with \emph{zero error probability} as long as $D\leq \Delta$, for an appropriate choice of $\Delta$. However, it is well-known that the minimum number of tests for these matrices satisfies\footnote{One should note that these bounds correspond to disjunct matrices. One can relax the disjunct property and yet achieve zero-error probability in conventional GT using the so-called separable matrices, which lead to the same asymptotic behavior as disjunct matrices~\cite{DH00,CH07}.}~\cite{N00}
\begin{align}\nonumber
\frac{\Delta^2}{2\log_2\Delta}\log_2n\  (1\!+\!o(1))\leq m\leq\Delta^2\log_2\e\log_2n\  (1\!+\!o(1)).
\end{align}
It is not difficult to show that $\Delta$ must be larger than $\lambda$ in order to have $\lim_{n\rightarrow\infty}P(D>\Delta)=0$. As a result, by requiring $P(\mathcal{E}|D\leq\Delta)=0$, one cannot obtain upper bounds on the number of tests for Poisson PGT that match the lower bound in Thm.~\ref{theorem:lower}. In order to overcome this problem, we instead used the less stringent condition $\lim_{n\rightarrow\infty}P(\mathcal{E}|D\leq\Delta)=0$ in Thms. \ref{thm_best} and \ref{thm_best2}, and employed the results of \cite{CD08} to obtain matching upper bounds on $m$. One should note that there exist other test constructions and decoding algorithms that may be used in conjunction with $\lim_{n\rightarrow\infty}P(\mathcal{E}|D\leq\Delta)=0$ to obtain matching upper bounds on $m$ for Poisson PGT (see for example~\cite{LCJS14,ABJ14,ABG14,SC15}); however, since an approach similar to the proof of Thms. \ref{thm_best} and \ref{thm_best2} can be used in these cases as well, we choose not to repeat these arguments and results. 

\begin{table}[t!]\centering
\caption{Lower and upper bounds on the minimum number of measurements $m$ using nonadaptive methods}
\begin{tabular}{|c|c|c|c|}
			\hline 
			$\!\!\!\!$Theorem $\!\!\!\!\!$&  Number of tests & Assumptions  \\ 
			\hline\hline
		
			Thm.~\ref{theorem:lower} & $m\!\!\geq\!\! (1\!-\!\epsilon)\lambda\log_2\!{n} (1\!-\!o(1))$ & $\lambda=o(n)$ \\\hline
			&  & $\lambda=o(n)$,\\
			Thm.~\ref{thm3}& $m\leq\e\lambda^{2+\epsilon}\log_2{n}  (1+o(1))$ & $\displaystyle\lim_{n\rightarrow\infty}\lambda=\infty$ \\\hline		
			&  & $\lambda=o(n)$,\\
			Thm.~\ref{thm4}& $m\!\!\leq\!\e\beta(n)^2\lambda^{2}\log_2\!{n}(1\!+\!o(1))$ & $\!\!0\!<\!\displaystyle\lim_{n\rightarrow\infty}\!\lambda\!<\!\infty\!\!$ \\\hline
						&  & $\lambda=o(n)$,\\			
			Thm.~\ref{thm_best}& $m\!\!\leq\!\!\frac{3}{\log_2 3}\lambda^{1\!+\!\epsilon}\log_2 n(1\!+\!o(1))$ & $\displaystyle\lim_{n\rightarrow\infty}\lambda=\infty$ \\\hline		
			&  & $\lambda=o(n)$,\\
			Thm.~\ref{thm_best2}& $m\!\!\leq\!\!\frac{3}{\log_2\! 3}\beta(n)\lambda\log_2 n(1\!+\!o(1))$ & $\!\!0\!<\!\displaystyle\lim_{n\rightarrow\infty}\!\lambda\!<\!\infty\!\!$ \\\hline			
						&  & $\lambda=o(n)$,\\			
			$\!\!$Thm.~\ref{thm1}$\!\!$& $\!\!\!m\!\!\leq\!\!2 \lambda^{\!1\!+\!\alpha}\!\!\left(\!\log n\!+\!c\beta\log^3\!\lambda\!\right)\!\!(1\!+\!o(1)\!)\!\!\!$ & $\displaystyle\lim_{n\rightarrow\infty}\lambda=\infty$ \\\hline		
			& $\!\!m\!\leq\!\!2{(\beta\lambda)}^{\!1\!+\!\gamma}(\!\log n\!+\!\tau\beta^2\!\log^2\!(\beta\lambda)\!)$  & $\lambda=o(n)$,\\
			$\!\!$Thm.~\ref{thm2}$\!\!$& $\  \  \  \  \  \  \  \  \  \  \  \  \  \   \  \  \  \  \  \  \  (1\!+\!o(1)\!)\!\!\!$ & $\!\!0\!<\!\displaystyle\lim_{n\rightarrow\infty}\!\lambda\!<\!\infty\!\!$ \\\hline

			\end{tabular}
			\label{table:nonadaptive}

\end{table}

\begin{table}[t!]\centering
\caption{Lower and upper bounds on the minimum expected number of measurements $\bar{m}$ using semi-adaptive methods}
\begin{tabular}{|c|c|c|c|}
			\hline 
			$\!\!\!\!$Theorem $\!\!\!\!\!$&  Number of tests & Assumptions  \\ 
			\hline\hline
		
			Thm.~\ref{theorem:lower_exp} & $\bar{m}>\lambda\log_2\!\frac{n}{\lambda}\  (1+o(1))\!\!$ & $\lambda=o(n)$ \\	
			 & $-\log_2\!\e\  \frac{\lambda^4}{n^2}$ &  \\\hline		
			$\!\!$Thm.~\ref{theorem:lower_exp}$\!\!$& $\bar{m}>\lambda\log_2\frac{n}{\lambda}(1+o(1))\!\!$ & $\!\!\lambda=o\left(\!(n^2\log_2n)^{\frac{1}{3}}\!\right)\!\!$ \\\hline		
			$\!\!$Thm.~\ref{theorem:upper_exp}$\!\!$& $\!\!\bar{m}\!\!\leq\!\!\frac{\e}{\log_2\!\e}{\lambda}\log_2\!\frac{n}{{\lambda}}(1\!+\!o(1))\!\!$ & $\lambda=o(n)$ \\\hline

			\end{tabular}
			\label{table:semiadaptive}

			\vspace*{-12pt}	
\end{table}

In the second part of our exposition (Sec.~\ref{sec:adaptive}), we focused on the family of semi-adaptive algorithms. These algorithms are performed in sequential stages, allowing to design new tests based on the outcome of previous tests in order to decrease the expected number of tests; in addition, in each stage the tests are designed and performed simultaneously, allowing parallel testing. In Sec.~\ref{sec:adaptive}, we used Huffman source coding to find a lower bound on the expected number of tests; in addition, we showed how Li's stage-wise algorithm~\cite{L62} developed for combinatorial GT can be modified for the Poisson PGT model. These lower and upper bounds are listed in Table~\ref{table:semiadaptive}.

Recent work in the area of group testing has almost exclusively focused on combinatorial GT. The results derived in this paper show that there exists a close connection between methods used for combinatorial GT and probabilistic GT. 

\appendices

\vspace*{-7pt}
\section{An information theoretic upper bound on the number of nonadaptive tests for Poisson PGT}\label{sec:method3}
Information theoretic approaches have been used in the study of \emph{combinatorial GT} problem by several authors~\cite{D04},~\cite{M78,MM80,AS12,SJ10}. In what follows, we apply these approaches to the Poisson PGT model in order to derive an upper bound on the minimum number of nonadaptive tests that satisfy~\eqref{condition1}. We assume that the test matrix is constructed probabilistically: the entries of the test matrix follow an i.i.d. Bernoulli$(p)$ distribution, such that each entry of $\mathbf{C}$ is equal to $1$ with probability $p$, and $0$ with probability $1-p$. For this construction method, we consider a maximum likelihood (ML) decoding procedure, which given the vector of test results and the test matrix reduces to:
\vspace*{-4pt}\begin{align}\nonumber
\hat{\mathcal{D}}&=\arg\max_{\mathcal{D}'} P(\mathbf{y},\mathbf{C}|\mathcal{D}')\\\label{eq:ML}
&=\arg\max_{\mathcal{D}'} P(\mathbf{y}|\mathbf{C},\mathcal{D}').
\vspace*{-4pt}\end{align}
Here, $P(\mathbf{y}|\mathbf{C},\mathcal{D}')$ denotes the conditional distribution of observing $\mathbf{y}$ given the test matrix $\mathbf{C}$ and set of defectives $\mathcal{D}'$. Note that the second equality holds since the test matrix is constructed independent of the set of defectives. 

The goal is to find the number of tests required to satisfy~\eqref{condition1}. We define the error event $\mathcal{E}'$ as the event that there exists a set of subjects $\mathcal{D}'\neq\mathcal{D}$ such that $P(\mathbf{y}|\mathbf{C},\mathcal{D}')\geq P(\mathbf{y}|\mathbf{C},\mathcal{D})$. It can be easily verified that $P(\mathcal{E})\leq P(\mathcal{E}')$. As a result, a number of tests that guarantees $\lim_{n\rightarrow\infty}P(\mathcal{E}')=0$ also guarantees~\eqref{condition1}. Given $D=d$, $1\leq d\leq n$, let $\mathcal{E}'_i$, $1\leq i\leq d$, denote the event that there exists a set of subjects with cardinality $d$, that differ from $\mathcal{D}$ in exactly $i$ items and is at least as likely as $\mathcal{D}$ to the decoder. Given these definitions, one has
\vspace*{-4pt}\begin{align}\nonumber
P(\mathcal{E}')=\mathbb{E}_D\left[P(\mathcal{E}'|D)\right]&=\sum_{d=1}^n c(n)\, \frac{{\lambda(n)}^d}{d!}\e^{-\lambda(n)} P\!\left(\cup_{i=1}^d\mathcal{E}'_i\right)\\
&\leq\sum_{d=1}^n \sum_{i=1}^d c(n)\, \frac{{\lambda(n)}^d}{d!}\e^{-\lambda(n)} P(\mathcal{E}_i')\label{Pe1},
\vspace*{-4pt}\end{align}
where the last inequality follows from the union bound.

At first glance, it may seem that a bound on $P(\mathcal{E}')$ may be obtained using an upper bound on $P(\mathcal{E}_i')$ for a \emph{fixed} value of $d$ (such as the bound presented in~\cite{AS12}), and subsequent averaging; however, there are two subtle, yet important issues that prohibit us from using this approach. First, in~\eqref{Pe1} the value of $d$, and hence $i$, may be as large as $n$. Since we are interested in the asymptotic regime where $n\rightarrow\infty$, a bound on $P(\mathcal{E}_i')$ should account for the growth of $d$ and $i$ with respect to $n$. Second, all known bounds on $P(\mathcal{E}_i')$ (see~\cite{AS12} and references therein) rely on a test matrix $\mathbf{C}$ with i.i.d. Bernoulli$(1/d)$ entries. However, in Poisson PGT, the true value of $d$ is unknown (more precisely, $D$ is a random variable) and cannot be used as a design parameter in a natural way. In order to overcome the aforementioned problems, we derive special functions that bound $P(\mathcal{E}_i')$ for different ranges of $d$, and in addition derive new bounds that do not rely on the value of $d$ as a design parameter. 

We start by observing that in~\cite{AS12}, it was shown that for $d=o(n)$, and for all $\rho, \, 0\leq \rho\leq1$, one has
\begin{subequations}\label{eq:PE}
\vspace*{-4pt}\begin{equation}
P(\mathcal{E}_i')\leq 2^{-m\left(E_o(\rho,i,d,n)-\frac{\rho\log{{n-d\choose i}{d\choose i}}}{m}\right)}
\vspace*{-4pt}\end{equation}
where the error exponent $E_o$ satisfies
\vspace*{-4pt}\begin{equation}\label{eq:Eo}
E_o(\rho,i,d,n)\!=-\!\log\!\!\!\!\sum_{Y\!\in\{0,1\}}\!\!\sum_{\  T_{2}}\!\!\left(\!\sum_{T_1}\!P(\mathbf{t}_1)P({y},\mathbf{t}_2|\mathbf{t}_1,\mathcal{D})^{\frac{1}{1+\rho}}\right)^{1+\rho}.
\vspace*{-4pt}
\end{equation}
\end{subequations}
\vspace{0.07in}

Here, we diverge slightly from the previously used notation and let $Y$ denote a random variable corresponding to the result of a \emph{single test} and let $y$ be a realization of $Y$. Let $(\mathcal{D}_1,\mathcal{D}_2)$ be a partition of $\mathcal{D}$ into disjoint sets with cardinalities $|\mathcal{D}_1|=i$ and $|\mathcal{D}_2|=d-i$, respectively. The vectors $T_1$ and $T_2$ are binary-valued row-vectors of length $i$ and $d-i$, indicating which subjects in $\mathcal{D}_1$ and $\mathcal{D}_2$ are present in a given test, respectively. Also, $\mathbf{t}_1$ and $\mathbf{t}_2$ are realizations of $T_1$ and $T_2$, respectively.

In order to prove the main results of this section, we need the following lemma.
\begin{lemma}\label{lemma:1}
Let $h(n):\mathbb{N}\mapsto{\mathbb{R}}^{+}$ be an increasing function of $n$ such that $\lim_{n\rightarrow\infty}h(n)=\infty$. Assume that each entry of the binary test matrix is an i.i.d. Bernoulli$(p)$ random variable, such that $\lceil h(n)\rceil\:p=o(n)$. Then $\forall i,d$ such that $1\leq i\leq d\leq \lceil h(n)\rceil$, and $\forall\rho$ such that $0<\rho<1$, one has the following bound on the error exponent:
\vspace*{-4pt}\begin{equation}\nonumber
E_o(\rho,i,d,n)\geq\rho(1-p)^dip\left(1-\frac{\rho}{2}\log^2(ip)+o(1)\right).
\vspace*{-4pt}\end{equation}
\end{lemma}
\begin{proof}
Given $h(n):\mathbb{N}\mapsto{\mathbb{R}}^{+}$, construct the test matrix $\mathbf{C}$ such that each entry follows an i.i.d. Bernoulli$(p)$ distribution, where $p=\lceil h(n)\rceil^{-(1+\epsilon')}$ for any fixed $\epsilon'$ such that $0<\epsilon'<1$. In order to prove this lemma, we use the results in \cite[Lemma VII.1]{AS12} and \cite[Lemma VII.2]{AS12} which state that $\forall i,d,n :1\leq i\leq d\leq n$,
\vspace*{-4pt}\begin{align}\label{eq1}
E_o(\rho,i,d,n)\geq&\rho I(T_1;Y|T_2)\!-\!\frac{\rho^2}{2}\!\max_{\rho:0\leq\rho<1}\left|\frac{\partial^2}{\partial \rho^2} E_o(\rho,i,d,n)\right|.
\vspace*{-4pt}\end{align}
If $\mathbb{E}_{T_1}[u_\rho\log u_\rho]$ is a non-increasing function of $\rho$, then
\vspace*{-4pt}\begin{equation}\label{eq:1616}
\left|\frac{\partial^2}{\partial \rho^2}E_o(\rho,i,d,n)\right|\leq\left| \sum_{T_2}\sum_{Y}g_\rho\  \mathbb{E}_{T_1}[u_\rho\log^2(u_{\rho})]\right|,
\vspace*{-4pt}\end{equation}
where we used the following notation 
\vspace*{-4pt}\begin{align}\nonumber
u_\rho&=\frac{P(y|\mathbf{t}_1,\mathbf{t}_2)^{1/(1+\rho)}}{\sum_{T_1}P(\mathbf{t}_1) P(y|\mathbf{t}_1,\mathbf{t}_2)^{1/(1+\rho)}},\\\nonumber
g_{\rho}&=\left({\sum_{T_1}P(\mathbf{t}_1) P(y,\mathbf{t}_2|\mathbf{t}_1)^{1/(1+\rho)}}\right)^{(1+\rho)}.
\vspace*{-4pt}\end{align}
In order to simplify the previous equations, we consider different realizations of $u_{\rho}$ for different values of $y$, $\mathbf{t}_2$ and $\mathbf{t}_1$. In particular, we consider four cases based on the realizations of the pair $(y,\mathbf{t}_2)$. For each case, we find $\mathbb{E}_{T_1}[u_\rho\log u_\rho]$ and show that this expectation is independent of $\rho$. In addition, when $\mathbb{E}_{T_1}[u_\rho\log u_\rho]\neq 0$, we find an expression for $g_\rho$.\\

\noindent\textbf{\emph{Case 1:}} Let $y=0$ and $\mathbf{t}_2=\mathbf{0}$. Then, we have
\vspace*{-4pt}\begin{equation}\nonumber
P(y=0|\mathbf{t}_1,\mathbf{t}_2=\mathbf{0})=\begin{cases}0&\  \text{if}\  \mathbf{t}_1\neq\mathbf{0}\\
1&\  \text{if}\  \mathbf{t}_1=\mathbf{0},
\end{cases}
\vspace*{-4pt}\end{equation}
which implies that 
\vspace*{-4pt}\begin{equation}\nonumber
u_\rho=\begin{cases}0&\  \text{if}\  \mathbf{t}_1\neq\mathbf{0}\\
\frac{1}{(1-p)^i}&\  \text{if}\  \mathbf{t}_1=\mathbf{0}.
\end{cases}
\vspace*{-4pt}\end{equation}
As a result,
\vspace*{-4pt}\begin{align}\nonumber
\mathbb{E}_{T_1}[u_\rho\log^2(u_{\rho})]&=\frac{(1-p)^i}{(1-p)^i}\log^2\left(\frac{1}{(1-p)^i}\right)\\\nonumber
&=\log^2\left(\frac{1}{(1-p)^i}\right)\\\label{eq:1717}
&=i^2\log^2(1-p).
\vspace*{-4pt}\end{align}
Since $P(y=0,\mathbf{t}_2=\mathbf{0}|\mathbf{t}_1)=P(y=0|\mathbf{t}_1,\mathbf{t}_2=\mathbf{0})P(\mathbf{t}_2=\mathbf{0})$, one has
\vspace*{-4pt}\begin{equation}\nonumber
P(y=0,\mathbf{t}_2=\mathbf{0}|\mathbf{t}_1)=\begin{cases}0&\  \text{if}\  \mathbf{t}_1\neq\mathbf{0}\\
P(\mathbf{t}_2=\mathbf{0})&\  \text{if}\  \mathbf{t}_1=\mathbf{0}.
\end{cases}
\vspace*{-4pt}\end{equation}
Consequently,
\vspace*{-4pt}\begin{align}\nonumber
g_{\rho}(y\!=\!0,\mathbf{t}_2\!=\!\mathbf{0})&\!=\!\left(\!{\sum_{T_1}\!P(\mathbf{t}_1) P(y\!=\!0,\mathbf{t}_2\!=\!\mathbf{0}|\mathbf{t}_1)^{1/(1\!+\!\rho)}}\right)^{\!(1\!+\!\rho)}\\\nonumber
&=\left({P(\mathbf{t}_1=\mathbf{0}) P(\mathbf{t}_2=\mathbf{0})^{1/(1+\rho)}}\right)^{(1+\rho)}\\\label{eq:18_2}
&=P(\mathbf{t}_1=\mathbf{0})^{(1+\rho)}P(\mathbf{t}_2=\mathbf{0}).
\vspace*{-4pt}\end{align}
Note that~\eqref{eq:18_2} implies that $g_{\rho}(y=0,\mathbf{t}_2=\mathbf{0})$ is a non-increasing function of $\rho$.

\noindent\textbf{\emph{Case 2:}} Let $y=1$ and $\mathbf{t}_2=\mathbf{0}$. Then, we have
\vspace*{-4pt}\begin{equation}\nonumber
P(y=1|\mathbf{t}_1,\mathbf{t}_2=\mathbf{0})=\begin{cases}1&\  \text{if}\  \mathbf{t}_1\neq\mathbf{0}\\
0&\  \text{if}\  \mathbf{t}_1=\mathbf{0},
\end{cases}
\vspace*{-4pt}\end{equation}
which implies that 
\vspace*{-4pt}\begin{equation}\nonumber
u_\rho=\begin{cases}\frac{1}{1-(1-p)^i}&\  \text{if}\  \mathbf{t}_1\neq0\\
0&\  \text{if}\  \mathbf{t}_1=0.
\end{cases}
\vspace*{-4pt}\end{equation}
As a result,
\vspace*{-4pt}\begin{align}\nonumber
\mathbb{E}_{T_1}[u_\rho\log^2(u_{\rho})]&=\frac{1-(1-p)^i}{1-(1-p)^i}\log^2\left(\frac{1}{1-(1-p)^i}\right)\\\label{eq:1818}
&=\log^2\left(1-(1-p)^i\right).
\vspace*{-4pt}\end{align}
Since $P(y=1,\mathbf{t}_2=\mathbf{0}|\mathbf{t}_1)=P(y=1|\mathbf{t}_1,\mathbf{t}_2=\mathbf{0})P(\mathbf{t}_2=\mathbf{0})$, one has
\vspace*{-4pt}\begin{equation}\nonumber
P(y=1,\mathbf{t}_2=\mathbf{0}|\mathbf{t}_1)=\begin{cases}P(\mathbf{t}_2=\mathbf{0})&\  \text{if}\  \mathbf{t}_1\neq\mathbf{0}\\
0&\  \text{if}\  \mathbf{t}_1=\mathbf{0}.
\end{cases}
\vspace*{-4pt}\end{equation}
Consequently,
\vspace*{-4pt}\begin{align}\nonumber
g_{\rho}(y\!=\!1,\mathbf{t}_2\!=\!\mathbf{0})&\!=\!\left(\!{\sum_{T_1}P(\mathbf{t}_1) P(y\!=\!1,\mathbf{t}_2\!=\!\mathbf{0}|\mathbf{t}_1)^{1/(1\!+\!\rho)}}\right)^{(1\!+\!\rho)}\\\nonumber
&=\left({P(\mathbf{t}_1\neq\mathbf{0}) P(\mathbf{t}_2=\mathbf{0})^{1/(1+\rho)}}\right)^{(1+\rho)}\\\label{eq:20_2}
&=P(\mathbf{t}_1\neq\mathbf{0})^{(1+\rho)}P(\mathbf{t}_2=\mathbf{0}).
\vspace*{-4pt}\end{align}
Note that~\eqref{eq:20_2} implies that $g_{\rho}(y=1,\mathbf{t}_2=\mathbf{0})$ is a non-increasing function of $\rho$.

\noindent\textbf{\emph{Case 3:}} Let $y=0$ and $\mathbf{t}_2\neq\mathbf{0}$. Then, we have
\vspace*{-4pt}\begin{equation}\nonumber
P(y=0|\mathbf{t}_1,\mathbf{t}_2\neq\mathbf{0})=0.
\vspace*{-4pt}\end{equation}
Consequently, $u_{\rho}=0$ and 
\vspace*{-4pt}\begin{equation}\label{eq:1919}
\mathbb{E}_{T_1}[u_\rho\log^2(u_{\rho})]=0.
\vspace*{-4pt}\end{equation}

\noindent\textbf{\emph{Case 4:}} Let $y=1$ and $\mathbf{t}_2\neq\mathbf{0}$. Then, we have
\vspace*{-4pt}\begin{equation}\nonumber
P(y=1|\mathbf{t}_1,\mathbf{t}_2\neq\mathbf{0})=1
\vspace*{-4pt}\end{equation}
Consequently $u_{\rho}=1$, and 
\vspace*{-4pt}\begin{equation}\label{eq:2020}
\mathbb{E}_{T_1}[u_\rho\log^2(u_{\rho})]=0.
\vspace*{-4pt}\end{equation}

In all these four cases, $\mathbb{E}_{T_1}[u_\rho\log u_\rho]$ is independent on $\rho$ and therefore a non-decreasing function of $\rho$. Substituting~\eqref{eq:1717}-\eqref{eq:2020} into~\eqref{eq:1616} gives
\vspace*{-4pt}\begin{align}\nonumber
&\left|\frac{\partial^2E_o(\rho,i,d,n)}{\partial \rho^2}\right|\leq\left| \sum_{T_2}\sum_{Y}g_\rho\  \mathbb{E}_{T_1}[u_\rho\log^2(u_{\rho})]\right|\\\nonumber
&=| g_\rho(y=0,\mathbf{t}_2=\mathbf{0})\  i^2\log^2(1-p)\\\nonumber
&\  \  \  \  \  \  \  \  \  \  \  \  \  \  \  \  \  \  \  \  \  \  +g_{\rho}(y=1,\mathbf{t}_2=\mathbf{0})\  \log^2\!\left({1\!-\!(1\!-\!p)^i}\right)\!|\\ \label{eq:2323}
&= g_\rho(y=0,\mathbf{t}_2=\mathbf{0})\  i^2\log^2(1-p)\\\nonumber
&\  \  \  \  \  \  \  \  \  \  \  \  \  \  \  \  \  \  \  \  \  \  +g_{\rho}(y=1,\mathbf{t}_2=\mathbf{0})\  \log^2\!\left({1\!-\!(1\!-\!p)^i}\right),
\vspace*{-4pt}\end{align}
where the last equality follows since $g_\rho$ is non-negative. Since $g_\rho(y=0,\mathbf{t}_2=\mathbf{0})$ and $g_\rho(y=1,\mathbf{t}_2=\mathbf{0})$ are non-increasing functions in $\rho$, we have
\vspace*{-4pt}\begin{align}\nonumber
\max_{\rho:0\leq\rho<1}g_\rho(y=0,\mathbf{t}_2=\mathbf{0})&=g_0(y=0,\mathbf{t}_2=\mathbf{0})\\\nonumber
&=P(\mathbf{t}_1=\mathbf{0})P(\mathbf{t}_2=\mathbf{0})\\\label{eq:2424}
&=(1-p)^d,
\vspace*{-4pt}\end{align}
and
\vspace*{-4pt}\begin{align}\nonumber
\max_{\rho:0\leq\rho<1}g_\rho(y=1,\mathbf{t}_2=\mathbf{0})&=g_0(y=1,\mathbf{t}_2=\mathbf{0})\\\nonumber
&=P(\mathbf{t}_1\neq\mathbf{0})P(\mathbf{t}_2=\mathbf{0})\\\nonumber
&=(1-p)^{d-i}(1-(1-p)^i)\\\nonumber
&=(1-p)^d\frac{1-(1-p)^i}{(1-p)^i}\\\nonumber
&=(1-p)^d\left((1-p)^{-i}-1\right)\\\label{eq:2525}
&=(1-p)^dpi(1+o(1)).
\vspace*{-4pt}\end{align}
Consequently, using \eqref{eq:2323}-\eqref{eq:2525}, it can be shown that
\vspace*{-4pt}\begin{align}\nonumber
&\max_{\rho:0\leq\rho<1}\left|\frac{\partial^2}{\partial \rho^2} E_o(\rho,i,d,n)\right|\\\nonumber
&\leq \max_{\rho:0\leq\rho<1}g_\rho(y=0,\mathbf{t}_2=\mathbf{0})\  i^2\log^2(1-p)\\\nonumber
&\  +\max_{\rho:0\leq\rho<1}g_{\rho}(y=1,\mathbf{t}_2=\mathbf{0})\  \log^2\!\left({1\!-\!(1\!-\!p)^i}\right)\\\nonumber
&= i^2(1-p)^d\log^2(1-p)\\\nonumber
&\  \  \  \  \  \  \  \  \  \  \  \  \  +(1-p)^{d-i}(1-(1-p)^i)\log^2\left({1-(1-p)^i}\right)\\\nonumber
&=(1\!-\!p)^d(i^2p^2(1\!+\!o(1))\!+\!ip\log^2\left({1\!-\!(1\!-\!p)^i}\right)(1\!+\!o(1)))\\\label{eq2}
&=(1-p)^dip\log^2(ip)\  (1+o(1)).
\vspace*{-4pt}\end{align}

Next, note that the mutual information in~\eqref{eq1} can be bounded according to 
\vspace*{-4pt}\begin{align}\nonumber
I(T_1;Y|T_2)&=H(Y|T_2)-H(Y|T_1,T_2)=\\\nonumber
&=H(Y|T_2)=(1-p)^{d-i}\  h\left((1-p)^i\right)\\\nonumber
&\geq (1-p)^{d-i}(1-p)^i\log(1-p)^{-i}\\\label{eq3}
&=(1-p)^d ip\:(1+o(1)).
\vspace*{-4pt}\end{align}
Substituting~\eqref{eq2} and~\eqref{eq3} into~\eqref{eq1} yields
\vspace*{-4pt}\begin{align}\nonumber
E_o(\rho,i,d,n)&\geq\rho I(T_1;Y|T_2)-\frac{\rho^2}{2}\max_{\rho:0\leq\rho<1}|E_o''(\rho,i,d,n)|\\\nonumber
&\geq\rho(1-p)^dip\left(1-\frac{\rho}{2}\log^2(ip)+o(1)\right).
\vspace*{-4pt}\end{align}
\end{proof}

We would like to point out that the previously proved lemma is a generalization of a lower bound on $E_o(\rho,i,d,n)$ from~\cite{AS12}, as the bound in~\cite{AS12} does not apply directly to the Poisson PGT model.

In order to find the number of tests that guarantee $P(\mathcal{E}')=o(1)$, we consider separately 
two asymptotic regimes for $\lambda(n)$: Theorem~\ref{thm1} presents the results for the asymptotic regime $\lambda(n)=o(n)$ and $\lim_{n\rightarrow\infty}\lambda(n)=\infty$; Similarly, Theorem~\ref{thm2} presents the results for the regime where $\lambda(n)=o(n)$, but $0<\lim_{n\rightarrow\infty}\lambda(n)<\infty$. Note that the case of constant $\lambda$ is covered by the latter scenario. 

\begin{theorem}\label{thm1}
Assume that $D$ follows the right-truncated Poisson distribution, with $\lambda(n)=o(n)$ and $\lim_{n\rightarrow\infty}\lambda(n)=\infty$. Construct a test matrix by choosing each entry according to a Bernoulli$(p)$ distribution, where $p=\lceil \lambda(n)^{(1+\epsilon)}\rceil^{-(1+\gamma)}$, for some fixed arbitrarily small scalars $\epsilon>0$ and $0<\gamma<1$. Under ML decoding, one can identify the set of defectives using $m=2 \lambda(n)^{1+\alpha}\left(\log n+c\:\beta(n)\log^3\lambda(n)\right)$ tests so that $\lim_{n\rightarrow\infty}P(\mathcal{E}')=0$, where $\beta(n)=\log^{(K)}n$ for some finite $K>1$, and $\alpha>0$ and $\tau>0$ are arbitrarily small fixed scalars. 
\end{theorem}

\begin{proof}
Since $\lambda(n)=o(n)$, there exists a fixed $\epsilon>0$ small enough such that $h(n)\triangleq\lambda(n)^{(1+\epsilon)}=o(n)$. Choose $p=\lceil h(n)\rceil^{-(1+\gamma)}$, 
for some $0<\gamma<1$. The probability of error given by formula~\eqref{Pe1} can be rewritten as $P(\mathcal{E}')\leq P_{e_1}+P_{e_2},$
where 
\vspace*{-4pt}\begin{align}\nonumber
P_{e_1}&=\sum_{d=1}^{\lceil h(n)\rceil}\sum_{i=1}^{d}c(n)\  \frac{\lambda(n)^d}{d!}\e^{-\lambda(n)}P(\mathcal{E}_i'),\\\nonumber
P_{e_2}&=\sum_{d=\lceil h(n)\rceil+1}^{n}\sum_{i=1}^{d}c(n)\  \frac{\lambda(n)^d}{d!}\e^{-\lambda(n)}P(\mathcal{E}_i').
\vspace*{-4pt}\end{align}
The idea is to bound these probabilities by finding a tight upper bound on $P(\mathcal{E}_i')$, independent of $i$ and $d$, for $1\leq d\leq\lceil h(n)\rceil$, while using the upper bound $P(\mathcal{E}_i')\leq 1$ for $\lceil h(n)\rceil+1\leq d\leq n$. Since $P(\mathcal{E}_i')\leq 1$, one has
\vspace*{-4pt}\begin{align}\nonumber
P_{e_2}&\leq\sum_{d=\left\lceil h(n)\right\rceil+1}^{n}c\:d\:\frac{\lambda(n)^d}{d!}\e^{-\lambda(n)}\\\nonumber
&\leq\sum_{d=\left\lceil h(n)\right\rceil+1}^{\infty}c\:d\:\frac{\lambda(n)^d}{d!}\e^{-\lambda(n)}\\\label{eq:app_rev1}
&=c\lambda(n)\sum_{d=\left\lceil \lambda(n)^{(1+\epsilon)}\right\rceil}^{\infty}\frac{\lambda(n)^d}{d!}\e^{-\lambda(n)}.
\vspace*{-4pt}\end{align}

The Chernoff bound for standard Poisson distributions ensures that for any $a\geq0$,
\vspace*{-4pt}\begin{equation}\label{chernoff}
\sum_{d=\lambda(n)+a}^{\infty}\!\!\!\!\frac{\lambda(n)^d}{d!}\e^{-\lambda(n)}\leq\exp\!\left(\!-(\lambda(n)+a)\log\frac{\lambda(n)+a}{\lambda(n)}+a\!\right).
\vspace*{-4pt}\end{equation}
Substituting $a\!=\!\lceil\lambda(n)^{(1+\epsilon)}\rceil-\lambda(n)$ for $\epsilon\!>\!0$ yields
\vspace*{-4pt}\begin{align}\nonumber
&\sum_{d=\lceil\lambda(n)^{(1+\epsilon)}\rceil}^{\infty}\!\!\!\!\!\frac{\lambda(n)^d}{d!}\e^{-\lambda(n)}\leq\exp\left(-\lceil\lambda(n)^{(1+\epsilon)}\rceil\log\lambda(n)^{\epsilon}\!+\!a\right)\\\label{app2:eq111}
&\  \  \  \  \  =\exp\left(-\epsilon\lceil\lambda(n)^{(1+\epsilon)}\rceil\log\lambda(n) \  (1+o(1))\right).
\vspace*{-4pt}\end{align}
Consequently, substituting~\eqref{app2:eq111} in~\eqref{eq:app_rev1} yields $P_{e_2}=o(1)$.

Now, the goal is to find the smallest value of $m$ such that $P_{e_1}=o(1)$. 
Since we have chosen $p=\lceil h(n)\rceil^{-(1+\gamma)}$, for some $0<\gamma<1$, using Lemma~\ref{lemma:1} one can show that $\forall i,d, \, 1\leq i\leq d\leq \lceil h(n)\rceil$ and $\forall\rho, \, 0<\rho<1$,
\vspace*{-4pt}\begin{align}\nonumber
&E_o(\rho,i,d,n) \geq \rho(1-p)^dip\left(1-\frac{\rho}{2}\log^2(ip)+o(1)\right) \\\nonumber
&\hspace{20pt}\geq \rho\:i \:p(1-p)^{\lceil h(n)\rceil}\left(1-\frac{\rho}{2}\log^2(p)+o(1)\right) \\\nonumber
&\hspace{20pt}= \rho\:i \:p\e^{-p\:\lceil h(n)\rceil}\left(1-\frac{\rho}{2}\log^2(p)\right) (1+o(1))\\\nonumber
&\hspace{20pt}\geq\frac{\rho\:i\:(1+o(1))}{{\lceil h(n)\rceil}^{1+\gamma}}\left(1-\frac{\rho}{2}(1+\gamma)^2\log^2(\lceil h(n)\rceil)\right)\\\nonumber
&\hspace{20pt}\geq\frac{\rho\:i\:(1+o(1))}{(h(n)\!+\!1)^{1+\gamma}}\!\left(1\!-\!\frac{\rho}{\!2}(1\!+\!\gamma)^2\log^2(h(n)\!+\!1)\right).
\vspace*{-4pt}\end{align}
By choosing $\rho=\frac{1}{(1+\gamma)^2\beta(n)\log^2(h(n)+1)}$, where $\beta(n)=\log^{(K)}n$ for some finite $K>1$, one arrives at
\vspace*{-4pt}\begin{equation}\nonumber
E_o(\rho,i,d,n)\geq\frac{\rho \:i}{(h(n)+1)^{1+\gamma}}\:(1+o(1)),
\vspace*{-4pt}\end{equation}
for any $i,d$ such that $1\leq i\leq d\leq\lceil h(n)\rceil$. In addition, using the inequality ${d\choose i}\leq{\left(\frac{d\e}{i}\right)}^i$, it can be easily shown that for $1\leq i,d\leq\left\lceil h(n)\right\rceil$, it holds that
\vspace*{-4pt}\begin{align}\nonumber
\log{{n-d\choose i}{d\choose i}}
&\leq i\log\left(\frac{d(n-d)\e^2}{i^2}\right)\\\nonumber
&\leq i\log\left(dn\e^2\right)\\\nonumber
&\leq 2 i\log n\:(1+o(1)).
\vspace*{-4pt}\end{align}
As a result, if 
\vspace*{-4pt}\begin{equation}\nonumber
m>\frac{2\rho \:i\log n}{\frac{\rho\:i}{(h(n)+1)^{1+\gamma}}}(1+o(1))=2{h(n)}^{1+\gamma}\log n\   (1+o(1)),
\vspace*{-4pt}\end{equation}
then $\left(E_o(\rho,i,d,n)-\frac{\rho\log{{n-d\choose i}{d\choose i}}}{m}\right)$ is positive. Therefore, using $m\geq(2+\delta(n))h(n)^{1+\gamma}\log n$ for any $\delta(n)>0$, we may write
\vspace*{-4pt}\begin{align}\nonumber
P(\mathcal{E}_i')&\leq 2^{-m\left(E_o(\rho,i,d,n)-\frac{\rho\log{{n-d\choose i}{d\choose i}}}{m}\right)}\\\nonumber
&\leq 2^{-m\rho\:i\left(\frac{1+o(1)}{(h(n)+1)^{1+\gamma}}-\frac{2\log n(1+o(1))}{m}\right)}\\\nonumber
&= 2^{-m\rho\:i\left(\frac{\delta(n)\log\!n{h(n)}^{1+\gamma}(1+o(1))}{mh(n)^{1+\gamma}}\right)}\\\nonumber
&= 2^{-\rho\:i\left(\delta(n)\log\!n(1+o(1))\right)}\\\nonumber
&\leq 2^{-\rho\left(\delta(n)\log\!n(1+o(1))\right)}\\\nonumber
&=2^{-\frac{\delta(n)\log\!n}{(1+\gamma)^2\beta(n)\log^2\! h(n)}(1+o(1))}\triangleq P_1(n).
\vspace*{-4pt}\end{align}
First, note that
\vspace*{-4pt}\begin{align}\nonumber
P_{e_1}&=\sum_{d=1}^{\lceil h(n)\rceil}\sum_{i=1}^{d}c(n)\  \frac{\lambda(n)^d}{d!}\e^{-\lambda(n)}P(\mathcal{E}_i')\\\nonumber
&\leq c(n)\,P_1(n)\sum_{d=1}^{\lceil h(n)\rceil}d\  \frac{\lambda(n)^d}{d!}\e^{-\lambda(n)}\\\nonumber
&\leq c(n)\, P_1(n)\sum_{d=1}^{\infty}d\  \frac{\lambda(n)^d}{d!}\e^{-\lambda(n)}\\\nonumber
&=c(n) P_1(n)\lambda(n)\\\nonumber
&=2^{-\frac{\delta(n)\log\!n}{(1+\gamma)^2\beta(n)\log^2 \!h(n)}+\log\lambda(n)(1+o(1))}.
\vspace*{-4pt}\end{align}
Therefore, if we choose
\vspace*{-4pt}\begin{equation}
\delta(n)=\epsilon(1+\epsilon)^2(1+\gamma)^2\beta(n)\log^3\lambda(n)/\log\! n,
\vspace*{-4pt}\end{equation}
one has
\vspace*{-4pt}\begin{align}\nonumber
P_{e_1}&\leq2^{-\frac{\delta(n)\log\!n}{(1+\gamma)^2\beta(n)\log^2 \!h(n)}+\log\lambda(n)(1+o(1))}\\\label{eq4}
&=2^{-\epsilon\log\lambda(n)(1+o(1))}=o(1).
\vspace*{-4pt}\end{align}

Consequently, the probability of error converges to zero, i.e., $P(\mathcal{E}')=o(1)$ if $m=(2+\delta(n))h(n)^{1+\gamma}\log n$. Substituting $h(n)=\lambda(n)^{1+\epsilon}$ in the previous expression, and performing some straightforward simplifications yields

\vspace*{-4pt}\begin{align}\nonumber
m= 2\lambda(n)^{1+\alpha}\left(\log n+\tau\:\beta(n)\log^3\lambda(n)\right)
\vspace*{-4pt}\end{align}
for some arbitrarily small fixed scalars $\alpha>0$ and $\tau>0$.
\end{proof}

\begin{theorem}\label{thm2}
Assume that $D$ follows the right-truncated Poisson distribution, with $\lambda(n)=o(n)$ and $0<\lim_{n\rightarrow\infty}\lambda(n)<\infty$. Let $\beta(n)=\log^{(K)}n$, for some finite $K>1$. Construct a test matrix by choosing each entry according to a Bernoulli$(p)$ distribution, where $p={\lceil\beta(n)\lambda(n)\rceil}^{-(1+\gamma)}$, for some fixed arbitrarily small scalar $\gamma>0$. Using ML decoding, one can identify defectives using $m= 2{(\beta(n)\:\lambda(n))}^{1+\gamma}(\log n+\tau\beta^2(n)\log^2(\beta(n)\lambda(n)))$ tests such that $\lim_{n\rightarrow\infty}P(\mathcal{E}')=0$, for scalars $\gamma>0$ and $\tau>0$. 
\end{theorem}

\begin{proof}
Let $h(n)\triangleq\beta(n)\:\lambda(n)$. Similar to the proof of Theorem~\ref{thm1}, we write $P(\mathcal{E}')\leq P_{e_1}+P_{e_2}$, where 
\vspace*{-4pt}\begin{align}\nonumber
P_{e_1}&=\sum_{d=1}^{\left\lceil h(n)\right\rceil}\sum_{i=1}^{d}\: c\  \frac{\lambda(n)^d}{d!}\e^{-\lambda(n)}P(\mathcal{E}_i'),\\\nonumber
P_{e_2}&=\sum_{\left\lceil h(n)\right\rceil+1}^{n}\sum_{i=1}^{d}\: c\  \frac{\lambda(n)^d}{d!}\e^{-\lambda(n)}P(\mathcal{E}_i').
\vspace*{-4pt}\end{align}
Since $P(\mathcal{E}_i')\leq 1$, one has
\vspace*{-4pt}\begin{align}\nonumber
P_{e_2}&\leq\sum_{d=\left\lceil h(n)\right\rceil+1}^{n}c\:d\:\frac{\lambda(n)^d}{d!}\e^{-\lambda(n)}\\\nonumber
&\leq\sum_{d=\left\lceil h(n)\right\rceil+1}^{\infty}c\:d\:\frac{\lambda(n)^d}{d!}\e^{-\lambda(n)}\\\label{eq:app_rev2}
&=c\lambda(n)\sum_{d=\left\lceil \beta(n)\:\lambda(n)\right\rceil}^{\infty}\frac{\lambda(n)^d}{d!}\e^{-\lambda(n)}.
\vspace*{-4pt}\end{align}
Since $\lim_{n\rightarrow\infty}\beta(n)=\infty$, there exists an integer $n'$ large enough such that $\beta(n)>1$, for all $n>n'$. Let $n>n'$ and substitute $a=\lceil \beta(n)\:\lambda(n)\rceil-\lambda(n)$ in~\eqref{chernoff}. Then,
\vspace*{-4pt}\begin{align}\nonumber
&\sum_{d=\lceil\beta(n)\:\lambda(n)\rceil}^{\infty}\frac{\lambda(n)^d}{d!}\e^{-\lambda(n)}\\\label{eq:lemma2b}
&\leq\exp\left(-\lceil\beta(n)\:\lambda(n)\rceil\log\beta(n)\!+\!\lceil \beta(n)\:\lambda(n)\rceil\!-\!\lambda(n)\right).
\vspace*{-4pt}\end{align}
As a result, substituting \eqref{eq:lemma2b} in \eqref{eq:app_rev2} yields $P_{e_2}=o(1)$.

Now, we find $m$ such that $p_{e_1}=o(1)$. The ideas behind the proof are similar to those described in the proof of Theorem~\ref{thm1}, except that in this case, we set $h(n)=\beta(n)\:\lambda(n)$ and $p=\lceil h(n)\rceil^{-(1+\gamma)}=\lceil \beta(n)\:\lambda(n)\rceil^{-(1+\gamma)}$, for some $0<\gamma<1$. As a result, $P(\mathcal{E}')=o(1)$, if $m=(2+\delta(n))h(n)^{1+\gamma}\log n$, where $\delta(n)=(1+\gamma)^2\beta^2(n)\log^2h(n)/\log\! n$. Substituting $h(n)=\beta(n)\:\lambda(n)$ in the previous expression, one arrives at $m= 2{(\beta(n)\:\lambda(n))}^{1+\gamma}(\log n+\tau\beta^2(n)\log^2(\beta(n)\lambda(n)))$, for some scalar $\tau>0$.
\end{proof}

\section*{Acknowledgment}
We would like to thank the anonymous reviewers for many useful comments and suggestions.
%

\ifCLASSOPTIONcaptionsoff
  \newpage
\fi


%
%
%




\end{document}